%% file: main_arXiv.tex
\documentclass{article}
\usepackage{algpseudocode}
\usepackage{algorithm}
\usepackage{hyperref}
\usepackage{graphicx}
\usepackage{ulem}
\input{./FileSettings/SetupMain}

\begin{document}

\input{Contents/TitlePage}
\input{Contents/MainContents}

\newpage
\bibliography{Contents/references}

\end{document}

%% file: FileSettings/SetupMain.tex
\input{FileSettings/MainPackages}
\input{FileSettings/Definitions}
\input{FileSettings/BibSettings}

%% file: FileSettings/MainPackages.tex
\usepackage{FileSettings/arxiv}
\usepackage[utf8]{inputenc} 
\usepackage[T1]{fontenc}    
\usepackage{url}            
\usepackage{booktabs}       
\usepackage{amsfonts}       
\usepackage{amsmath}
\usepackage{nicefrac}       
\usepackage{microtype}      
\usepackage{lipsum}		
\usepackage{graphicx}
\usepackage{listings}
\usepackage[table]{xcolor}
\usepackage{xcolor}
\usepackage[most]{tcolorbox} 
\usepackage[framemethod=tikz]{mdframed}
\usepackage[labelfont=bf]{caption}
\usepackage{hyperref}
\usepackage{tabularx}

%% file: FileSettings/Definitions.tex
\definecolor{codegreen}{rgb}{0,0.6,0}
\definecolor{codegray}{rgb}{0.5,0.5,0.5}
\definecolor{codepurple}{rgb}{0.58,0,0.82}
\definecolor{backcolour}{rgb}{0.95,0.95,0.92}

\lstdefinestyle{mystyle}{
  commentstyle=\color{codegreen},
  keywordstyle=\color{magenta},
  numberstyle=\tiny\color{codegray},
  stringstyle=\color{codepurple},
  basicstyle=\ttfamily\footnotesize,
  breakatwhitespace=false,         
  breaklines=true,                 
  captionpos=b,                    
  keepspaces=true,                 
  numbers=left,                    
  numbersep=10pt,                  
  showspaces=false,                
  showstringspaces=false,
  showtabs=false,                  
  tabsize=2,
  xleftmargin=5pt,
  aboveskip=5pt,
  belowskip=0pt
}

\definecolor{light-gray}{gray}{0.95}

\surroundwithmdframed[
  hidealllines=true,
  backgroundcolor=light-gray,
  innerleftmargin=0pt,
  skipabove=10pt,
  skipbelow=20pt]{lstlisting}

\definecolor{HeaderGray}{gray}{0.85}	
\definecolor{Gray}{gray}{0.9}
\definecolor{LightCyan}{rgb}{0.78,0.97,0.97}

\DeclareMathOperator{\sgn}{sgn}
\DeclareMathOperator{\rand}{rand_U}

\newcommand{\subfigref}[2]{\hyperref[{#1}]{\ref*{#1}#2}}

%% file: Contents/TitlePage.tex
\title{Probabilistic Computing Optimization of Complex Spin-Glass Topologies}

\author{%
    \href{https://orcid.org/0009-0007-9273-4480}{
    \includegraphics[scale=0.06]{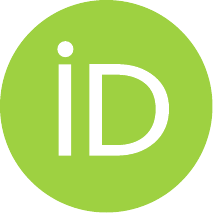}\hspace{0.5mm}\textbf{Fredrik Hasselgren}}\textsuperscript{1,2,*} \and
    \href{https://orcid.org/0009-0007-1800-3241}{
    \includegraphics[scale=0.06]{Figures/Misc/orcid.pdf}\hspace{0.5mm}\textbf{Max Al-Hasso}}\textsuperscript{1,$\dagger$} \and
    \href{https://orcid.org/0000-0003-1795-871X}{
    \includegraphics[scale=0.06]{Figures/Misc/orcid.pdf}\hspace{0.5mm}\textbf{Amy Searle}\textsuperscript{1}} \and
    \href{https://scholar.google.com/citations?user=OdXhja4AAAAJ}{
    \hspace{0.5mm}\textbf{Joseph Tindall}}\textsuperscript{3} \and
    \href{https://orcid.org/0000-0002-7192-0256}{
    \includegraphics[scale=0.06]{Figures/Misc/orcid.pdf}\hspace{0.5mm}\textbf{Marko von der Leyen}}\textsuperscript{1,4}
}

\date{}

\maketitle

\begin{center}
    \textsuperscript{1} Quantum Dice Ltd, 264 Banbury Road, Oxford, OX2 7DY, United Kingdom \\
    \textsuperscript{2} Mathematical Institute, University of Oxford, Oxford, OX2 6GG, United Kingdom \\
    \textsuperscript{3} Center for Computational Quantum Physics, Flatiron Institute, New York, New York 10010, USA \\
    \textsuperscript{4} Department of Physics, University of Oxford, Oxford, OX1 3PU, United Kingdom \\
    \vspace{0.25cm}
    \textsuperscript{*}\texttt{fredrik.hasselgren@keble.ox.ac.uk}
    \textsuperscript{$\dagger$}\texttt{max.al-hasso@quantum-dice.com}
\end{center}

\vspace{1cm}

\input{Contents/Abstract}
\keywords{Probabilistic computing, spin-glass, Max-Cut, QUBO}

%% file: Contents/Abstract.tex
\begin{abstract}
Spin glass systems as lattices of disordered magnets with random interactions have important implications within the theory of magnetization and applications to a wide-range of hard combinatorial optimization problems. Nevertheless, despite sustained efforts, algorithms that attain both high accuracy and efficiency remain elusive. Due to their topologies being low-$k$-partite such systems are well suited to a probabilistic computing (PC) approach using probabilistic bits (P-bits). Here we present complex spin glass topologies solved on a simulated PC realization of an Ising machine. First, we considered a number of three dimensional Edwards-Anderson cubic spin-glasses randomly generated as well as found in the literature as a benchmark. Second, biclique topologies were identified as a likely candidate for a comparative advantage compared to other state-of-the-art techniques, with a range of sizes simulated. We find that the number of iterations necessary to find solutions of a given quality has constant scaling with system size past a saturation point if one assumes perfect parallelization of the hardware. Therefore a PC architecture can trade the computational depth of other methods for parallelized width by connecting a number of P-bits that scales linearly in system size. This constant scaling is shown to persist across a number of solution qualities, up to a certain limit beyond which resource constraints limited further investigation. The saturation point varies between topologies and qualities and becomes exponentially hard in the limit of finding the ground truth. Furthermore we demonstrate that our PC architecture can solve spin-glass topologies to the same quality as the most advanced quantum annealer in minutes, making modest assumptions about their implementation on hardware. 

\end{abstract}

%% file: Contents/MainContents.tex
\textbf{}\input{Contents/Introduction}
\input{Contents/SpinGlass}

\input{Contents/Biclique}
\input{Contents/Discussion}
\input{Contents/Acknowledgments}

%% file: Contents/Introduction.tex
\section{Introduction}

\newcommand{\jt}[1]{{\color{red}{#1}}}

\subsection{Problem formulation and related work}

\subsubsection{Spin-glasses and Max-Cut}
The spin glass model has received sustained interest \cite{binder_spin_1986, mezard_spin_1987} due to the fact that a broad spectrum of discrete optimization problems can be mapped to the task of finding the model's low energy states \cite{barahona_computational_1982, farhi_quantum_2001, battaglia_optimization_2005, lucas_ising_2014, zhang_mapping_2025}. The added difficulty in finding these ground states lies in the frustration which arises from conflicting interactions within the lattice. We follow \cite{lucas_ising_2014, mott_solving_2017, king_quantum_2021, abel_quantum-field-theoretic_2021, mohseni_ising_2022, mezard_spin_1987, perdomo-ortiz_finding_2012, dridi_prime_2017, jiang_quantum_2018, criado_qade_2022, paszynski_portfolio_2021, zhang_computational_2025} in restricting our Hamiltonian to a quadratic form of Ising spin interactions of $N$ sites. These interactions are encoded in the entries $J_{ij}$ of an $N \times N$ symmetric matrix $J$, leading to the Hamiltonian
\begin{equation}
    \mathcal{H} = \sum_{i < j} J_{ij} \sigma_i^z \sigma_j^z.
    \label{eq:Hamiltonain}
\end{equation}

One important example of a spin glass is the Edwards-Anderson (EA) model \cite{mezard_spin_1987, edwards_theory_1975, sherrington_solvable_1975, binder_spin_1986, young_spin_2002, boettcher_physics_2024, zhang_computational_2025} of only nearest-neighbor interactions, whose topology is that of a $D$-dimensional hypercubic lattice. This topology is found in various areas of physics and is well studied in the literature. The problem of finding EA model ground-states in $D>2$ was proven to be NP-hard \cite{barahona_computational_1982, istrail_statistical_2000, hartmann_ground_2011, garey_computers_1979, arora_computational_2009}, and no known algorithm has been found that solves the problem in polynomial time \cite{boettcher_physics_2024, hastings_monte_1970, kirkpatrick_optimization_1983}. In the case of a three-dimensional lattice, it was proven \cite{zhang_computational_2020} that algorithmic complexity is lower bounded by $2^{N/3}$ for $N \to \infty$ given the \textit{exponential time hypothesis} \cite{impagliazzo_complexity_1999}. In their recent work, Ref. ~\cite{zhang_computational_2025} points out that algorithms or purpose-built hardware is still permitted a scaling of $2^{N/b}$, more efficient for $N < b^3$, finding $b \approx 10^3$ using D-Wave's Advantage $4.1$ Annealer \cite{johnson_quantum_2011, dickson_thermally_2013, ohkuwa_reverse_2018, cao_speedup_2021, king_coherent_2022, king_quantum_2023, bernaschi_quantum_2024} and classical postprocessing. With the D-Wave annealer a ground truth for specific instances of $N= 678, 958, 1312, 2084$ was found, along with a likely contender for $N= 5627$.

The ground truths in Ref. \cite{zhang_computational_2025} were found using a digital cooling technique employed on many low energy states, motivating the usefulness of finding low energy approximations even in contexts where consistently finding the ground truth is unfeasible. In general, finding low energy (comparatively optimized) states can still bring enormous benefit even without determining the ground truth optimal state. 

A spin-glass graph is considered sparse if the number of instantiated connections as a fraction of all possible connections approaches zero as the lattice size goes to infinity and dense otherwise. It is well-established in the literature that dense spin-glass graphs are a harder category to solve than sparse graphs \cite{zbinden_embedding_2020, hamerly_experimental_2019, zeng_performance_2024, johnson_quantum_2011, bunyk_architectural_2014, boixo_evidence_2014, boev_quantum-inspired_2023, okada_improving_2019}. As such, for the present work we consider two types of graphs: cubic sparse topologies and biclique dense topologies. We begin by considering randomly generated cubic lattices as well as the 3D topologies found in Ref. \cite{zhang_computational_2025} in \autoref{seq:3D}, solving for low energy states using a probabilistic computing architecture. We then move on to the biclique topology which is related to neural networks via Restricted Boltzmann Machines (RBM) and has a density approaching $1/2$ for the full bi-partite connections considered here. This topology is of special interest since its dense connectivity poses difficulties to current best-in-class simulation approaches. For instance, D-Wave's latest quantum annealer shows noticeably increasing errors with system size \cite{king_beyond-classical_2025} compared to other, sparser, topologies whilst lattice-specific tensor networks --- which have been used to overturn a number of quantum advantage claims \cite{tindall_dynamics_2025, tindall_efficient_2024, rudolph_simulating_2025, begusic_fast_2024} --- cannot efficiently encode the extensive number of degrees of freedom associated with each vertex. Therefore, we also investigate the effect of this connectivity on the PC efficacy for comparison in \autoref{seq:biclique}.

Finding the ground state energy of spin glass models can be recast as a weighted max-cut problem. This graph theoretical NP-hard problem sets out to determine a partition of a graph such that the number of edges crossing the partition is maximized. As stated by Ref. \cite{garey_simplified_1976} the exponential time hypothesis claims that the max-cut problem cannot be solved in $2^{o(N)}$, forming an upper bound on algorithm efficiencies. The fastest algorithm for solving the max-cut problem was proposed by Williams \cite{williams_new_2004} in 2004 and has a polynomial bounded time complexity of $\mathcal{O}^*(1.74^N)$ that requires exponential space \cite{lalovic_exact_2024}. As such one of the "open problems around exact algorithms" outlined by Woeginger is whether a polynomial space algorithm that runs faster than $\mathcal{O}^*(2^N)$ exists \cite{woeginger_open_2008}. To this end, many classical algorithms for solving max-cut problems have been proposed, each with its own scaling and suitability to different subclasses of graphs. In the present work we demonstrate an Ising machine heuristic solver, finding solutions within some tolerance of the ground truth in exchange for large computational savings.

\subsubsection{Ising machines}

Ising machines are solvers implemented via specialized hardware or software that aim to find ground states of the Ising model \cite{mohseni_ising_2022}.  This category of solver thus encompasses CPU-searching \cite{leleu_scaling_2021, kowalsky_3-regular_2022, mandra_deceptive_2018}, Monte Carlo algorithms \cite{mandra_deceptive_2018}, dynamical chaotic algorithm  \cite{leleu_scaling_2021}, dynamical oscillators \cite{hamerly_experimental_2019} , simulated annealing hardware \cite{aramon_physics-inspired_2019, kowalsky_3-regular_2022, cai_power-efficient_2020, patel_ising_2020, albash_demonstration_2018}, and quantum annealers \cite{hamerly_experimental_2019, kowalsky_3-regular_2022, albash_demonstration_2018, mandra_deceptive_2018} to name a few. RBMs and thus the probabilistic computing architecture presented here employing Boltzmann machines fall into this category. Due to their varying nature, different realizations of Ising machines come with their own benefits and drawbacks compared to the others. For this project we will not attempt to make comparisons to the entirety of this broad ecosystem but instead we will focus our discussion to the probabilistic classical annealing based machinery presented here and its closest relative: quantum annealing.

In 2020, an RBM was mapped onto a reconfigurable Field Programmable Gate Array (FPGA) and used to solve the dense max-cut problem where the coupling matrix was randomly initialized with all couplings set to $0$ or $1$ with an equal likelihood \cite{patel_ising_2020}. This problem is simpler than the weighted max-cut formulated above and the problem sizes considered only went up to $N=200$, but we expect significant carry-over between the two formulations. This hardware-accelerated RBM was benchmarked against the D-Wave 2000Q Quantum Adiabatic Computer \cite{blais_operation_2000, johnson_quantum_2011} and the Optical Coherent Ising Machine \cite{inagaki_coherent_2016, mcmahon_fully_2016} achieving asymptotic scaling either similar or better for both \cite{patel_ising_2020}. The authors report a $10^7$x and $150$x time to solution improvement for their max-cut problem compared to the Quantum Adiabatic Computer and Coherent Ising Machine. Furthermore a $10-20$x improvement over classical CPU algorithms like those deployed on probabilistic computing was demonstrated. These results are highly motivating for the present investigation as our computational approaches have a common root in Boltzmann machines and our biclique topology matches their RBM strongly.


\subsection{Probabilistic computing architecture}
\subsubsection{Probabilistic computing fundamentals}
A Boltzmann machine is a stochastic Ising model made up of a graph of binary nodes with a total energy derived from the system's Hamiltonian. In the context of probabilistic computing a Boltzmann machine's binary units are implemented as a graph made up of probabilistic bits (P-bits) and their connectivity. The P-bits are defined by their bias \textit{I} used to compute their output \textit{m} that will always be in the set $\{ 0,1 \}$ according to some probability distribution $\{ P(0), P(1)\}$. We follow \cite{camsari_stochastic_2017} in specifying to the case of
\begin{equation}
    m = \sgn \{\sigma(I) + \rand[-1,+1]\},
    \label{eq:bias}
\end{equation}
where $\sigma$ is the sigmoid function, $\rand[-1,+1]$ is a uniform random number on the interval $[-1,+1]$ and we apply a $\frac{m}{2} +\frac{1}{2}$ transformation, to convert bipolar outputs to binary values.

The VCBM architecture operates by calculating the difference in energy arising from a given P-bit reading $0$ or $1$ and using that difference as the updated bias. In the quadratic Ising spin glass Hamiltonian $\mathcal{H}$ this cancels out most terms in Eq. \eqref{eq:Hamiltonain} leaving only
\begin{equation}
\Delta E_i(\textbf{m})
\;\equiv\;
\mathcal{H}(\textbf{m}^{(i \leftarrow 1)})-\mathcal{H}(\textbf{m}^{(i \leftarrow 0)})
\;=\;
\sum_{j\in\mathcal N(i)} J_{ij}\, m_j \;
\;=\; I_i .
\label{eq:deltaE}
\end{equation}
where $\mathcal{N}(i)$ are the neighbors of the \textit{i}-th lattice site and the system's configuration is $\textbf{m}=(m_1, m_2, \dots, m_N)$. Equations \eqref{eq:bias} and \eqref{eq:deltaE} together evolve the connected P-bits to a steady state distribution defined by eigenvector with eigenvalue $+1$ of the corresponding Markov matrix \cite{chowdhury_full-stack_2023, camsari_stochastic_2017}. Therefore when $J$ is symmetric the steady state distribution of the graph will be described by the Boltzmann law \cite{mascagni_simulated_1990}
\begin{equation}
    p(E(\mathbf{m})) = \frac{e^{- \beta E(\mathbf{m})}}{Z}.
\end{equation}
Where here $\beta$ is the temperature of the system, $E(\textbf{m})$ is the energy as a function of the current state vector and $Z$ is the partition function. Thus repeated cycles of P-bit bias calculations approximates the Boltzmann distribution without tackling the intractable problem of finding \textit{Z} exactly, irrespective of the order the P-bits were updated \cite{chowdhury_full-stack_2023}. Thus, the PC architecture can be applied to problems where the ground state represents the solution to some optimization problem such that the system will, with high likelihood, evolve towards this state over time. The specifics of how to approach the ground state depends on the algorithm deployed, which come with different advantages and drawbacks when it comes to finding low energy states without getting stuck in local minima. 

In any regard, this Boltzmann machine's convergence toward low-energy states is not determined solely by the number of P-bits in the system. This is directly analogous to statistical physics, where the equilibration time of a gas is controlled by microscopic scattering rates and correlation lengths rather than the absolute number of particles. In other words, the relevant timescale is set by the local dynamics that govern mixing of the distribution, not by system size itself. Similarly, in a PC Boltzmann machine, convergence is dictated by the spectral gap of the corresponding Markov transition matrix and by the effective interaction range of the P-bits; hence, enlarging the system does not inherently imply a slower relaxation to the steady-state Boltzmann distribution.

\subsubsection{Update groups}
One key concept in probabilistic computing is the notion of an \textit{update group}. In the process described above it is clear that for a given cycle there will be P-bits who, due to being disconnected, have no causal influence on each other. These P-bits form an update group such that all groups make a vertex partition of the P-bit graph where there are no internal edges in the groups. In the case of a universal vertex the size of the update group can be one, and in the case of an edgeless graph there is one update group with all \textit{N} nodes. The edgeless graph can clearly not perform any of the probabilistic computation outlined above, and therefore the smallest number of update groups capable of computation is two.

The P-bits in an update group can safely be updated in parallel, leading to a computational speedup in purpose-built hardware. Holding the size of the lattice constant, the larger the size of the update groups, either on average or at minimum, the larger this speedup will be. For a general graph, finding these update groups is the NP-hard graph coloring problem, and so it will not always be feasible for large complex graphs. In these instances a greedy coloring heuristic can be employed — i.e., vertices are colored sequentially by always assigning each vertex the smallest available color not used by its already-colored neighbors. Although this method does not guarantee an optimal solution, it runs in polynomial time and is guaranteed to use at most \(\Delta + 1\) colors, where \(\Delta\) is the graph’s maximum degree — making it a practical and efficient fallback when exact coloring is computationally infeasible \cite{borowiecki_computational_2018}.

In the context of the Edwards-Anderson model of spin-glasses, one of the advantages is then the bi-partite nature of the update groups. For balanced lattices this guarantees the largest speedup possible as it has both the optimal number and size of update groups. This motivates the application of the PC protocol to solving the spin-glass optimization performed in this paper. Furthermore, in instances of a common neighbor between elements of an update group a further speedup can be achieved by sampling the neighbor only once and re-using the value for the bias calculation of all connected nodes. The size of this improvement will then depend on the average number of shared neighbors in the update groups. In the spin-glass topologies with repeating boundary conditions such as those considered here this is clearly also maximized since having only two update groups means that every neighbor is shared between elements of the update groups. In the case of a biclique topology this effect is realized to the largest extent possible as all neighbors are shared between all elements of an update group. In hardware realizations this can then be incorporated to maximize the efficiency of PC architecture on spin-glasses by taking the union of the sets of neighbors of all vertices in an update group and sampling each neighbor once, storing the values for subsequent bias calculations.

These results translate into concrete hardware guidance: maximize bipartite structure (or minimize the chromatic number) to enlarge update groups, and implement shared-neighbor sampling so that each neighbor is read once per group and reused across connected P-bits. Together, these strategies reduce per-iteration work and improve effective update bandwidth, directly informing P-bit data paths, memory access patterns, and scheduler design in probabilistic computing hardware.

\subsection{Methods}
\subsubsection{Optimization algorithm}
\label{seq:optimization}
\begin{figure}[H]
    \centering
    \includegraphics[width=1.0\linewidth]{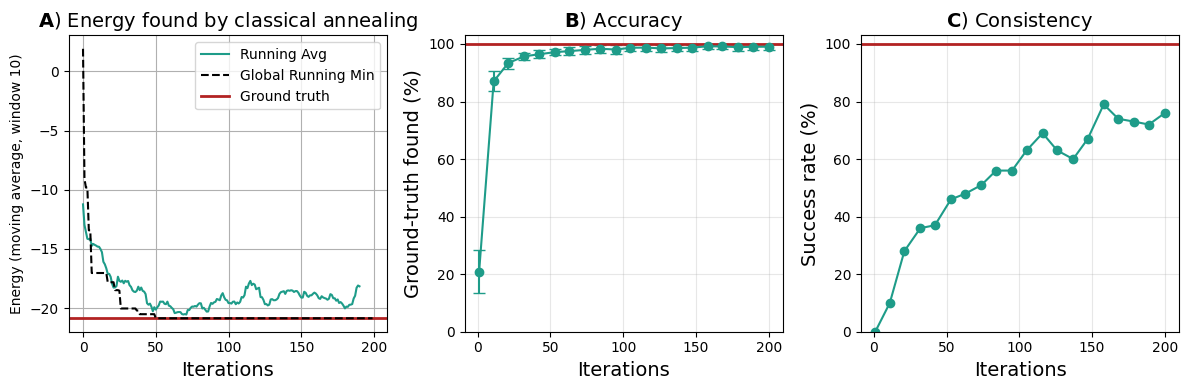}
    \caption{Performance of the PC solver. \textbf{A)} Energies found by the simulated annealing algorithm both with the running average with a window of size $10$ for smoothness and with the running lowest energy found. \textbf{B)} Quality defined as the fraction of the ground truth found on average which converges to $1$ as the iterations increases. \textbf{C)} Success rate of all the runs considered for various iterations, generally increasing with more iterations.}
    \label{fig:PCVCBM}
\end{figure}

The optimization algorithm employed on the PC architecture in this paper is classical annealing (CA). We chose this as it is a competitive alternative heuristic algorithm to quantum annealing, despite not possessing the typical traits heralded as important in quantum annealing, such as tunneling. In this protocol a temperature variable $\beta(i)$ is introduced where $i$ is the iteration number. The temperature is generally a monotonically decreasing function whose inverse $1/\beta(i)$ modulates the bias of all the P-bits in the system. Thus a higher temperature brings the biases closer to zero, making their outputs more random and reflecting a higher entropy in the system. This allows the PC to explore a larger area of the configuration space. As the temperature is decreased, the P-bit outputs become more and more deterministic with the decreasing entropy, lessening the rate of displacement within the phase space. The performance of CA on finding the lowest energy of a cubic lattice of 27 sites can be seen in \autoref{fig:PCVCBM} \textbf{A)}.

The idea for this heuristic algorithm is then to converge on low energy states by slowly turning down the temperature according to some schedule, tuning the PC towards the ground state. Additionally a step duration alongside step-size can be implemented in the schedule such that the PC has time to converge on a good configuration at the given temperature before it decreases. The hope is to avoid local minima by sufficiently slowly turning down the temperature to always be able to overcome energy barriers separating local and global minima. In essence any temperature schedule can be implemented but for this project a linear temperature schedule with optional step durations was used.

\subsubsection{Classical solvers}
\label{seq:classical}

For the purposes of establishing the ground truth of the biclique topologies studied in this paper, we utilize the \textit{MQLib} library developed for Ref. \cite{dunning_what_2018}. This library contains implementations of $37$ different classical solving algorithms solving for the ground state of our spin glass formulated as a Quadratic Unconstrained Binary Optimization (QUBO) problem matrix. All algorithms were run to solve the given instances, with the ground truth being taken as the consensus of the algorithms. Due to the limited sizes of the topologies considered, a high level of agreement between the algorithms was achieved, indicating a likely contender for the ground truth. Nevertheless, for the largest sizes considered here there is the possibility of inaccuracy in the classical solutions found.

\subsubsection{Scaling analysis}

The PC architecture employing CA is expected to have a likelihood to converge on the ground truth that increases with iteration number \textit{i}. Therefore, in the limit of large \textit{i} and a large number of replica instances the solver is expected to find the minima, or a close approximation of it, of any finite system. Furthermore, the convergence behavior follows the characteristic profile of many heuristic optimization algorithms, where the largest improvements occur during the early stages of the search. Subsequent gains diminish rapidly, exhibiting an approximately exponential decay in progress before flattening into a regime of slow incremental improvement. This saturation-like convergence pattern is well-documented in both stochastic optimization and machine learning, reflecting the rapidly decreasing probability of escaping near-optimal basins once the majority of the search space has been efficiently explored \cite{viering_shape_2023}. The PC solution coalescing on the ground truth can be seen in \autoref{fig:PCVCBM} \textbf{B)} for the solution quality and \textbf{C)} for the success rate of a given run.

Due to this expected behavior, it is possible to identify the identify the minimal number of iterations $i_{\rm min}$ such that a given replica instance has a $s$ success likelihood of converging on a solution within the percentage $p$ of the actual ground truth value. In the case of $p=1$ this is clearly a test for ground truth finding, but due to the nature of heuristic solvers and the shape of its improvement over iterations choosing a smaller value better showcases its performance. With these considerations \textbf{Algorithm 1} was deployed in this project for determining $i_{\rm min}$ to benchmark the solver's performance. In the case that one wants to look at a series of instances of increasing system size, this algorithm can be chained sequentially, setting $i_{\rm start}$ to the found $i$ of the previous instance as generally larger instances will require a larger computational budget.
\begin{algorithm}[H]
\caption{Minimal $i$ passing the tolerance test with linear $\beta$ schedule}
\begin{algorithmic}[1]
\Require Instance $\mathcal{I}$, ground truth $G$, start $i_{\mathrm{start}}$, step $i_{\mathrm{step}}$, trials $t$, threshold factor $p$, success target $p_{\mathrm{succ}}$, $\beta_i$, $\beta_f$
\State $i \gets i_{\mathrm{start}}$;\quad $\theta \gets p\,G$;\quad $m_{\max}\gets \mathrm{round}\!\big(t(1-p_{\mathrm{succ}})\big)$ \Comment{threshold and max allowed fails}
\While{true}
  \State $\delta\beta \gets (\beta_f-\beta_i)/i$;\quad $m\gets 0$ \Comment{linear $\beta$ step per iteration; reset fail counter}
  \For{$j=1$ \textbf{to} $t$}
    \State $v \gets \textsc{PC}(\mathcal{I},\, i,\, \beta_i,\, \beta_f,\, \delta\beta)$ \Comment{run solver for $i$ iterations with linear schedule}
    \If{$v>\theta$}
       \State $m\gets m+1$ \Comment{count a fail for this trial}
       \If{$m>m_{\max}$} \State \textbf{break} \Comment{early exit: too many fails at this $i$} \EndIf
    \EndIf
  \EndFor
  \If{$m\le m_{\max}$} \State \Return $i$ \Comment{passed: all $t$ trials without exceeding fail cap}
  \Else \State $i\gets i+i_{\mathrm{step}}$ \Comment{increase iteration budget and retry}
  \EndIf
\EndWhile
\end{algorithmic}
\end{algorithm}

%% file: Contents/SpinGlass.tex
\section{Cubic three-dimensional Ising spin glass model}
\label{seq:3D}

Here we consider a cubic three-dimensional Ising spin glass model with each vertex possessing six neighbors and periodic boundary conditions. Such a topology is sparsely connected, and so the majority of vertex pairs do not share an edge. We will refer to such pairs as \textit{virtually connected}, meaning that their interaction is mediated indirectly via intermediate vertices. Indeed starting from a given lattice point and moving a distance away the number of virtual connections necessary to bridge two nodes increases linearly. As reported in \cite{zhang_computational_2025} the \textit{trivial-non-trivial} picture of spin-glasses \cite{krzakala_spin_2000, palassini_nature_2000} predicts an energy landscape containing a unique ground state alongside low energy basins defined by a cluster of low energy states within some Hamming distance of each other. The number of such basins rises exponentially as one considers a larger energy threshold compared to the ground truth. Using the PC architecture we implement the optimization techniques discussed in \autoref{seq:optimization} to find low energy states of the lattice. To investigate the scaling of the iterations of the PC necessary to find solutions of a given quality \textbf{Algorithm 1} was deployed on cubic lattices of size $3^3, 4^3, \dots, 10^3$. Similarly, for a point of performance comparison,  the lattices from \cite{zhang_computational_2025} were solved using their established ground truths as benchmarks.

\subsection{Iterations scaling for random cubic lattices}

\begin{figure}
    \centering
    \includegraphics[width=1.0\linewidth]{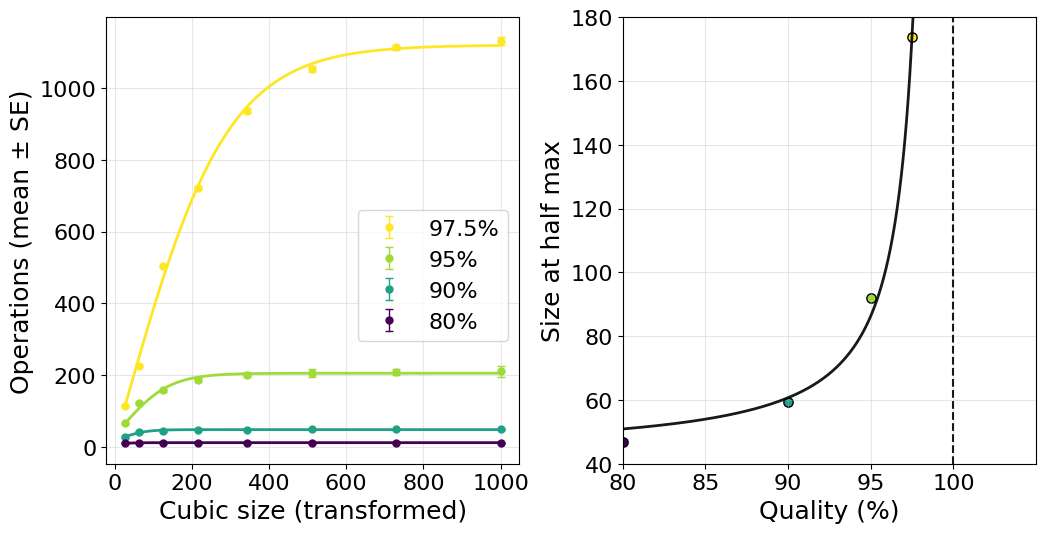}
    \caption{\textbf{Left: }The minimum number of PC iterations necessary to pass \textbf{Algorithm 1.} and generate spin-glass solutions to various accuracy percentages for 20 trials at a $90\%$ success rate for cubic lattices of sizes $N=2^3, 3^3, \dots, 10^3$. These values are curve-fitted to a logistic function \eqref{eq:logistic}. \textbf{Right: }The lattice size at of the half-maximum plotted alongside a exponential fit with a asymptote at $100\%$ quality.}
    \label{fig:cubic-random}
\end{figure}

For this testing 31 datasets of randomly generated cubic lattices were solved according to \textbf{Algorithm 1}. The resultant complexity scaling is shown in \autoref{fig:cubic-random} and demonstrates a constant time complexity with size, persisting over different solution qualities. This means that for a constant solution quality, increasing the size of the cubic lattice does not incur an additional iteration cost beyond the point of saturation. This is assuming perfect parallelization in that the number of P-bits updated in parallel for each iteration scales as $N/2$. This shows that beyond the total number of P-bit updates increasing with our system size, our time-to-solution (TTS) will not increase with system size as long as parallelization is maintained. From \autoref{fig:cubic-random}, the number of iterations required to attain higher solution qualities clearly increases exponentially which is as expected since the problem converges on its normal NP-hard formulation as the quality approaches $100\%$.

The complexity scaling indicated by \autoref{fig:cubic-random} appears to continue across the different solution qualities tested here, but the plausible continuation of the trend is hard to prove. Indeed there could be a higher-class scaling with a very small pre-factor present that goes undetected for the sizes considered here. Due to the limitations on ground-truth generation for these problems, continuing our analysis into larger problem sizes would require a different reference energy and is therefore a point of possible future investigation.

The iteration scalings shown in \autoref{fig:cubic-random} and \autoref{fig:bicliqueScaling} appear to follow a logistic function shifted such that it begins at the origin, as the number of iterations should tend to zero alongside the size of the state-space. With this in mind one can curve-fit the scalings to this shape, noting that the the statistical variation from solving random problem instances will lead to an imperfect fit. Here we fitted the data to an anchored logistic
\begin{equation}
    f(x) \;=\; L \left( \frac{1 + e^{-k x_c}}{1 + e^{-k(x - x_c)}} \;-\; e^{-k x_c} \right), 
    \label{eq:logistic}
\end{equation}

with steepness parameter $k$, y-asymptote $L$ and an inflection point given at $x=x_c$.

\subsection{High-quality solutions}

A range of lattices with $N= 678, 958, 1312, 2084, 5627$ from Ref. \cite{zhang_computational_2025} were tailored to be compatible with running on the D-Wave annealer architecture. This process involves a randomly generated lattice dimerized at each point, and then subjected to a rewiring of the edges to fit the Pegasus architecture \cite{zhang_computational_2025, boothby_next-generation_2020}. As a result the PC implementation of these lattices required $4$ update groups for the smallest instance and $5$ for the others. Furthermore, to match the methodology of \cite{zhang_computational_2025} we consider the solution quality metric of the inverse effective temperature defined as the inverse residual energy per site
\begin{equation}
    b_{\mathrm{eff}} \approx \frac{N}{E-E_0}
    \label{b}
\end{equation}
for a found energy $E$ compared to the ground truth $E_0$. If one follows the post-processing technique outlined in \cite{zhang_computational_2025} of digital cooling one can reach a true inverse temperature of $b=(2\log2)\frac{E_0}{N}b_{\mathrm{eff}}$ for a sufficiently large ensemble of low-energy states.

\begin{figure}
    \centering
    \includegraphics[width=1.0\linewidth]{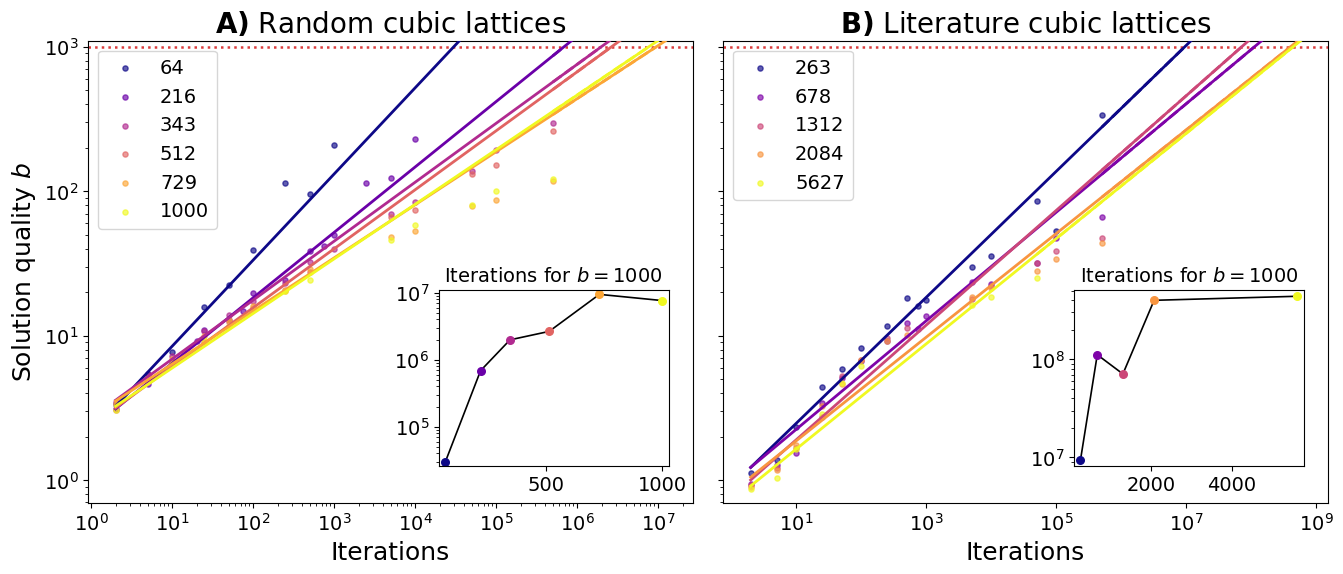}
    \caption{Power law scaling of the \textit{b} value attained for cubic lattices color-coded by size, solved by a range of different iterations. Alongside the data are least-squares power-laws where each datapoint was weighted by the inverse variance in log-pace arising from the error propagation of the standard error of the heuristic solver. \textbf{A) }for randomly generated cubic lattices each with its own ground-truth across $n=31$ different datasets. \textbf{B) } Scaling for engineered 3D cubic lattices from Ref. \cite{zhang_computational_2025} for which a ground-truth value is known, averaged across $n=12$ different runs. \textbf{Insets: }are the extrapolated number of iterations necessary to reach $b=10^3$ over system size for random and literature cubic lattices.}
    \label{fig:power-law}
\end{figure}

To reach high-quality solutions at $b \approx 10^3$, the effect of solving the randomly generated cubic lattices with various number of iterations was tested. This yielded a power-law relationship between the number of iterations used and the \textit{b}-value achieved. For smaller cubic topologies, attaining $b\approx 10^3-10^5$ was possible even in the simulation, with the largest values probably resulting from the solver routinely finding the actual ground truth. For larger topologies reaching this quality was infeasible, but a clear power-law behavior allows extrapolation. This power law scaling is shown in \autoref{fig:power-law} \textbf{A)} and \textbf{C)}.

The power law extrapolation shows that larger cubic topologies require a larger number of iterations to attain the same high-quality results as smaller lattices. For small values of \textit{b} the same relationship is not demonstrated. Therefore, these results match those from \autoref{fig:cubic-random} which demonstrated that whilst the number of operations required for a given quality saturates over time and ceases to increase with system size, this saturation point lies higher for higher quality solutions. The saturation lattice site appears to grow exponentially with quality which coincides with no saturation point being possible in the limit of $100\%$ accuracy.

As discussed and shown in \autoref{fig:cubic-random} we expect the number of iterations necessary to attain a quality of $b\approx10^3$ to saturate past a certain lattice site. From \autoref{fig:power-law} \textbf{B)} and its inset it is not clear if that size has been reached due to the significant error bounds incurred when extrapolating the power laws by orders of magnitude. Whilst this dataset, limited by the computational bounds of our simulation, cannot confirm this, we believe that this data is indicative of a saturation point as one would expect from \autoref{fig:cubic-random}.

The saturation behavior is further evidenced by the fact that the random and literature lattices solved in \autoref{fig:power-law} both indicate the plausible saturation despite being different in their topology, with the former being a perfect lattice with two update groups and the latter being a representation of a cubic lattice in the Pegasus architecture topology resulting in either four or five update groups. This explains the vertical shift between the two collections of datasets, as a higher number of update groups should yield a multiplicative factor in the amount of iterations required to reach a certain quality, which would manifest in precisely such a shift in the logarithmic plot.

%% file: Contents/Biclique.tex
\section{Biclique topologies}
\label{seq:biclique}

Infinite dimensional spin glass topologies coming in the form of bicliques have proven especially challenging for both D-Wave quantum annealers and MPS tensor network methods \cite{king_beyond-classical_2025}. For a probabilistic computing architecture this topology comes with the benefit that it is bipartite and there will always be only two update groups, the theoretically optimal number. To test the PC performance on biclique topologies, $n=31$ datasets of random bicliques with sizes ranging from $N=20$ to $N=400$ were generated with random couplings chosen from $[-1,1]$. These datasets were solved by the CA protocol, tracking the lowest number of iterations necessary to solve to within $p$ of the classical ground truth to $s$ success likelihood. This shows number of iterations required to produce a certain quality of solutions across system sizes.

\subsection{Iterations scaling with system size}

\begin{figure}
    \centering
    \includegraphics[width=1.0\linewidth]{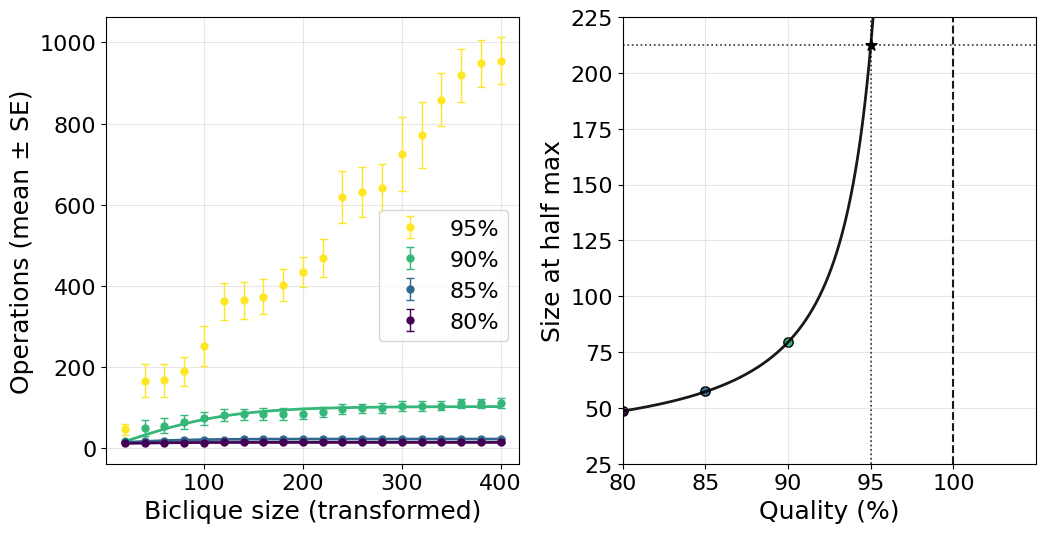}
    \caption{\textbf{Left: }The minimum number of PC iterations necessary to pass \textbf{Algorithm 1.} to various accuracy percentages for \textit{20} trials at a $90\%$ success rate for bicliques of increasing number of sites. These values are curve-fitted to a logistic function \eqref{eq:logistic}. \textbf{Right: }The lattice size at the half-maximum plotted alongside a exponential fit with a asymptote at $100\%$ quality alongside an extrapolation to what size one expects for $95\%$.}
    \label{fig:bicliqueScaling}
\end{figure}

To investigate the scaling of the iterations necessary to solve the spin glass to a certain quality, different values of $p$ were used for $s=0.90$ at 20 trials per test. This resulted in \autoref{fig:bicliqueScaling} showing a constant scaling between the number of iterations required and the system size past a certain saturation point similarly to the relations shown in \autoref{fig:cubic-random}. The trend becomes less clear for higher solution qualities, making it hard to say with certainty that the trend is quality-independent. Furthermore, as can be seen in right hand side of \autoref{fig:bicliqueScaling}, the lattice size at which the inflection point of the curve is reached (at half the height of the asymptote) increases exponentially with solution quality. This is likely the reason why we cannot see the iterations saturating for the highest qualities as this indicates that the points at which the saturation becomes visible may lie outside of our considered lattice sizes. Indeed, extrapolating the curve-fit of these data points indicates that the half-maximum of the $95\%$-quality solution is within our range at close to $N=200$. However, since the flattening of the logistic function only becomes apparent at a higher multiple of this size it is as expected that we are not able to see it within our range.

Theoretical considerations comport with these results as well. Since the complexity goes to $2^N$ as the accuracy approaches $100\%$, the size at which one sees the saturation needs to scale accordingly. If the size at which the asymptote becomes apparent scaled in an order lower than this complexity scaling then the gradient of the line would pick up this scaling instead. This would result in scenarios where a PC could solve one lattice to a certain quality trivially whilst failing to solve the next size up for a runtime of weeks which seems unlikely given our results, and how we expect the navigation of the energy landscape to look.

Therefore, for a limited set of system sizes, as is the case here due to ground-truth limitations, we expect there to be qualities whose saturation is indemonstrable by our numerical experiments. We believe that is the behavior shown in \autoref{fig:bicliqueScaling}, but acknowledge that there could be a small-prefactor higher-order scaling present but unseen in our considered ranges. Overall, this indicates that the complexity scaling for a perfectly parallel PC solving biclique lattices is plausibly $\mathcal{O}(1)$ just like for cubic lattices. The increased computational difficulty in demonstrating this is likely a consequence of the heightened complexity of the biclique topologies compared to the sparse cubic lattices.

\begin{figure}
    \centering
    \includegraphics[width=0.8\linewidth]{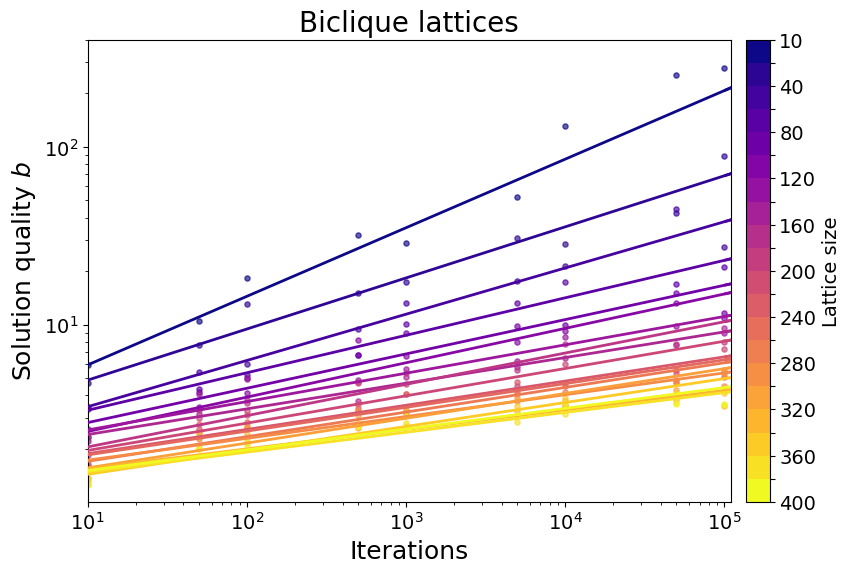}
    \caption{Power-law scaling of the \textit{b} value attained for biclique lattices solved over a range of iterations, alongside their weighted curve-fits as in \autoref{fig:power-law}. This data had significantly higher error in the curve-fit and so extrapolation is omitted for accuracy.}
    \label{fig:biclique-power}
\end{figure}

Lastly, we performed similar power-law calculations on the bicliques as was performed for \autoref{fig:power-law}. The PC was run with iterations ranging from $10^2$ to $10^5$ and the resultant weighted curve-fits are shown in \autoref{fig:biclique-power}. The accuracy of these fits was generally lower than for the cubic topologies which is possibly an artifact of the density of these graphs compared to the sparser graphs. As such, accurate extrapolation of these lines is not possible and we instead focus our analysis on their apparent behavior. As can be seen in \autoref{fig:biclique-power} the power-laws appear to adhere to two general trends. Firstly the y-intercept and slope decrease with increasing lattice size. This entails that larger lattices are generally harder to solve for the sizes considered here. The second trend is that the lines appear to plausibly converge with larger lattice sizes. This would mean that for sufficiently large lattices, the rise in iterations for the same quality with size flattens out and saturates which is precisely the behavior found in \autoref{fig:bicliqueScaling}. 
We believe that this data and the earlier results are mutually re-enforcing in their indication of the saturation of the minimal iteration count.

%% file: Contents/Discussion.tex
\section{Discussion}
\subsection{Hardware implementation}

To fully leverage the benefits of our PC approach, a hardware implementation will have to be built. Ongoing work indicates that photonics-based P-bit implementations will be able to operate at clock speeds above $250 \,\mathrm{MHz}$. The efficient nature of the bias calculation for cubic lattices and its suitability for FPGA implementations means a computational overhead on the order of tens of steps. This number is the expected bottleneck imposed by the hardware amounting to $n_{\mathrm{cycles}} = \log_2(N)+\mathcal{O}(10)$ where the logarithmic term arises from the scaling of the adder tree algorithm and the other term denotes miscellaneous hardware processes of constant duration. Furthermore, hardware considerations indicate that in a QUBO with couplings sampled from a discrete distribution, the $ \log_2(N)$ addition could be done in one step making the process overall $\mathcal{O}(1)$.

This would mean that the estimated $10^8$ iterations necessary to solve cubic lattices to a quality of $b\approx10^3$ would take on the order of minutes on a fully parallelized P-bit computer with current estimates. The precise number of clock cycles for each iteration can be determined after hardware implementation and will include the P-bit updating and bias calculation. If any of these assumptions are not met, for example, if the clock speed, P-bit update rate, or number of logical P-bits are lower than assumed, this would only result in a constant multiplicative increase in the total time to solution. For example if the probabilistic computer has half the logical P-bits needed for one spin-glass update group, twice as many clock cycles will need to pass to perform the same number of iterations. Thus we expect our time to solution to scale in hardware as $\mathcal{O}(1)$ for fully parallelized machines or alternatively as $\mathcal{O}(N)$ when this is not feasible.
 
\subsection{General QUBO solver}

The PC spin-glass solver presented here can be straightforwardly extended to solving general QUBOs that may not be bipartite like the problems hitherto considered. The key requirement for this is a method by which the update-groups of the problem instance can be computed. As discussed, this is by itself an NP-hard problem for a most generally formulated instance. If a cheap greedy coloring heuristic is deployed then update groups can be generated cheaply without necessarily finding the optimal solution which would dampen the efficacy of the implementation by some constant multiplicative factor. Nevertheless, the same considerations that make the PC suited to generating quality solutions to the spin-glass ground state problem should carry over to  more general QUBO problems. This effect is strengthened the fewer colors necessary to color the graph, which could inform the choice of when to deploy this architecture compared to other mechanisms for solving QUBOs.

As demonstrated in \autoref{fig:power-law} \textbf{A)} and \textbf{C)}, the number of update groups does not affect the power-law complexity exhibited by our PC when solving spin-glasses. Instead, sub-optimal numbers of update groups incur a multiplicative factor for the number of iterations required to attain a certain quality solution, since it takes on average more iterations to perform a full sweep of the system. This will then lead to an increased time-to-solution when compared to an optimal bipartite graph, but we will still be able to take advantage of the complexity scaling demonstrated in this paper. Indeed in the limit of the number of update groups equaling the number of nodes (for a completely connected graphical representation of a QUBO), this factor will reflect the fact that our PC will no longer have any capacity for parallelization in its p-bit updates. The speedup arising from only reading the state of common neighbors once can still arise in dense graphs, but also goes to zero as the graph approaches full connectivity (when all nodes share all neighbors). Therefore, we expect the current implementation to perform better on more sparse QUBO's, falling in line with other Ising machines. Nevertheless as demonstrated in \cite{al-hasso_probabilistic_2025}, the PC Ising machine is well-suited to solving some problems even with \textit{N} update groups for \textit{N} bits.

\section{Conclusion}

\subsection{High-quality solutions}
The authors of \cite{zhang_computational_2025} conclude that for $N < b^3$ annealing devices provide the most efficient way to produce high-quality solutions from which their digital cooling protocol can infer the ground state. As they point out, exact numerical algorithms like branch-and-cut \cite{mezard_information_2009} like those reported in \cite{palassini_low-energy_2003} can reach $b\approx10^2$, whilst their quantum annealer could reach $b\approx10^3$. In this paper we have demonstrated that even with the inherent limitations of just simulating a PC we were able to reach solutions consistently as high as $b\approx10^5$ for our smallest cubic systems. For the largest lattice considered here and in \cite{zhang_computational_2025} we found a scaling of $b \approx 7.07 \, i^{0.37}$ where \textit{i} is the number of iterations of the PC, with $R^2=0.98$. Due to the variance in our random datasets solved by a heuristic solver we cannot conclude that these estimates are accurate, only that they give an indication for the possible performance of a future hardware PC implementation. If this power-law continues past the range considered here we can extrapolate it to the data-points $(b=10^2, i\approx10^5), (b=10^3, i\approx10^8), \, \& \, (b=10^4,i\approx10^{11})$. We agree with \cite{zhang_computational_2025} in that there is no reason to expect a fundamental limit on $b$ for a large enough spin-glass and so high-quality computation comes down to the scaling necessary to reach high-$b$ solutions.

We expect better scalings than those demonstrated in \autoref{fig:power-law} to be possible with more sophisticated solving techniques than CA and with more experimentation performed to determine the optimal temperature schedule for solving a specific instance. For the experimentation performed here, a linear annealing schedule starting at $\beta=0.35$ and ending on $\beta=3.5$ was deployed. We have no reason to believe that this is the universally optimal schedule, or indeed that the optimal schedule is invariant between instances. Therefore we expect future experimentation optimizing the schedule to be able to bring down the scaling factor of our complexity, likely significantly decreasing the number of iterations for estimated above. Furthermore, different PC solving methods like parallel tampering and others were not part of the present investigation but could likewise yield an lower scaling pre-factor than those presented here. In \cite{searle_virtually_2025}, parallel tampering was shown to solve certain spin-glasses better on PC architecture than classical annealing, indicating a possible speedup arising from considering the solving method more carefully.

The likely ability for PC hardware to "cool" the system to comparable temperatures in its simulation invites the question of what quantum behavior contributes to solving the optimization problems in \cite{zhang_computational_2025} that classical purpose-built hardware cannot achieve. The answer to this question likely needs the hardware implementation of the theoretical PC presented here to be fully benchmarked. Our testing indicates that even for the largest system sizes, temperatures well beyond $10^{-3}$ is possible for cubic topologies by running the PC for longer or by optimizing the solving protocol. Whilst $b$-values significantly higher than those reached here and in \cite{zhang_computational_2025} are likely possible in future works utilizing a quantum annealer, the simulations in this work suggest it is unclear that current quantum-annealing devices can reach qualities unobtainable for specialized classical hardware.

\subsection{Scaling}

In this project we provided strong evidence that the amount of iterations necessary for PC architecture to reach a certain quality of solution to be constant in the size of the problem considered, assuming full parallelization. This reflects how in the idealized spin-glass topologies with two update-groups each iteration is $N/2$ P-bit updates wide through parallelization. Thus even with the number of iterations held constant, the actual number of P-bit updates scales with increasing system size, which is required for the Boltzmann machine to anneal towards the ground truth to a constant quality. This allows the PC to trade computational depth for parallelized width to an extent determined by the number of P-bits in each update group.

If we extend our analysis to more connected topologies, we would as discussed need to use a graph-coloring algorithm to determine the update-groups, yielding more update groups. In the limit of a fully connected graph we would need $N$ update groups. On average, this means we would need \textit{N} iterations to perform a full sweep of the graph, previously achieved in a constant two iterations. Therefore, we expect the worst-case scenario for the most highly connected topologies to have a scaling in their number of iterations required to reach a constant quality of solutions over system size of $\mathcal{O}(N)$, yielding the same scaling in the time-to-solution. Taken together, the observed constant-iteration scaling under ideal parallelization suggests that, for the spin-glass topologies studied here, purpose-built probabilistic hardware could achieve competitive time-to-solution without quantum resources—thereby challenging the necessity of quantum annealers for these instances, pending hardware realization and broader benchmarking. 

\subsection{Sparsification}

Recent work like \cite{aadit_massively_2022, sajeeb_scalable_2025} have developed sparsification techniques to transfer worse-performing dense Ising machines into better-performing sparse Ising machines. This work has been done specifically in the context of probabilistic computing to successfully out-compete other Ising machine implementations \cite{nikhar_all--all_2024, chowdhury_pushing_2025}. The sparsification algorithm proposed in \cite{sajeeb_scalable_2025} allows the specification of the maximum number of neighbors a site can have, and leaves the update groups intact. This protocol comes at the cost of introducing significantly more P-bits and having to optimize for the strength of the ferromagnetic copy edge deployed in the technique, adding to our approximation inaccuracy. Nevertheless, the efficacy of this technique raises the question of whether one could use sparsification make QUBO/spin-glass problems easier to solve since our PC TTS is insensitive to system size increases in an idealized scenario.

In the present work we have demonstrated a constant iterations scaling in system size, along with an overall harder time solving more connected biclique topologies than sparser cubic topologies. If a sparsification technique was deployed on the biclique lattice with the maximum resultant number of neighbors being set to six, our findings indicate that the complexity of solving this problem would now be that of the cubic spin-glasses, which would seemingly be lower. As discussed, the additional system size penalty incurred by the technique would not result in additional iterations required, only hardware P-bits implemented. This would then increase the feasibility of solving larger biclique spin-glasses if one is able to implement larger numbers of P-bits instead. As the authors of \cite{sajeeb_scalable_2025} report, this overhead is lessened when only approximate solutions are sought, such as for our heuristic PC architecture. At this stage, we cannot determine if the scaling advantages of transforming a dense graph into a sparse graph through sparsification outweighs the incurred overhead. Instead, we highlight this as a promising area of future research into solving dense spin-glass topologies on probabilistic computers.

%% file: Contents/Acknowledgments.tex
\section*{Author Contributions}
\label{Contributions}
F.H. wrote the implementation, performed the simulations, and authored the manuscript. M.A., A.J.S, J.T. and M.L. provided continual feedback and resources shaping the development of the project. M.L. and M.A. supervised the project. All authors contributed to editing of the manuscript. All authors contributed substantially to discussions on the results and ideas contained therein. 

\section*{Acknowledgments}
\label{acknowledgments}
J.T. is grateful for ongoing support through the Flatiron Institute, a division of the Simons Foundation. F.H. is thankful for continued support from the Mathematical Institute Scholarship with Jane Street Graduate Scholarship.

%% file: main_arXiv.bbl
\begin{thebibliography}{10}
\urlstyle{rm}
\expandafter\ifx\csname url\endcsname\relax
  \def\url#1{\texttt{#1}}\fi
\expandafter\ifx\csname urlprefix\endcsname\relax\def\urlprefix{URL }\fi
\expandafter\ifx\csname doiprefix\endcsname\relax\def\doiprefix{DOI: }\fi
\providecommand{\bibinfo}[2]{#2}
\providecommand{\eprint}[2][]{\url{#2}}

\bibitem{binder_spin_1986}
\bibinfo{author}{Binder, K.} \& \bibinfo{author}{Young, A.~P.}
\newblock \bibinfo{journal}{\bibinfo{title}{Spin glasses: {Experimental} facts, theoretical concepts, and open questions}}.
\newblock {\emph{\JournalTitle{Reviews of Modern Physics}}} \textbf{\bibinfo{volume}{58}}, \bibinfo{pages}{801--976}, \doiprefix\url{10.1103/RevModPhys.58.801} (\bibinfo{year}{1986}).
\newblock \bibinfo{note}{Publisher: APS ADS Bibcode: 1986RvMP...58..801B}.

\bibitem{mezard_spin_1987}
\bibinfo{author}{Mezard, M.}, \bibinfo{author}{Parisi, G.} \& \bibinfo{author}{Virasoro, M.~A.}
\newblock \emph{\bibinfo{title}{Spin glass theory and beyond}}.
\newblock No.~\bibinfo{number}{9} in \bibinfo{series}{World {Scientific} lecture notes in physics} (\bibinfo{publisher}{World scientific}, \bibinfo{address}{Teaneck, NJ, USA}, \bibinfo{year}{1987}).

\bibitem{barahona_computational_1982}
\bibinfo{author}{Barahona, F.}
\newblock \bibinfo{journal}{\bibinfo{title}{On the computational complexity of {Ising} spin glass models}}.
\newblock {\emph{\JournalTitle{Journal of Physics A: Mathematical and General}}} \textbf{\bibinfo{volume}{15}}, \bibinfo{pages}{3241--3253}, \doiprefix\url{10.1088/0305-4470/15/10/028} (\bibinfo{year}{1982}).

\bibitem{farhi_quantum_2001}
\bibinfo{author}{Farhi, E.} \emph{et~al.}
\newblock \bibinfo{journal}{\bibinfo{title}{A {Quantum} {Adiabatic} {Evolution} {Algorithm} {Applied} to {Random} {Instances} of an {NP}-{Complete} {Problem}}}.
\newblock {\emph{\JournalTitle{Science}}} \textbf{\bibinfo{volume}{292}}, \bibinfo{pages}{472--475}, \doiprefix\url{10.1126/science.1057726} (\bibinfo{year}{2001}).

\bibitem{battaglia_optimization_2005}
\bibinfo{author}{Battaglia, D.~A.}, \bibinfo{author}{Santoro, G.~E.} \& \bibinfo{author}{Tosatti, E.}
\newblock \bibinfo{journal}{\bibinfo{title}{Optimization by quantum annealing: {Lessons} from hard satisfiability problems}}.
\newblock {\emph{\JournalTitle{Physical Review E}}} \textbf{\bibinfo{volume}{71}}, \bibinfo{pages}{066707}, \doiprefix\url{10.1103/PhysRevE.71.066707} (\bibinfo{year}{2005}).
\newblock \bibinfo{note}{Publisher: American Physical Society}.

\bibitem{lucas_ising_2014}
\bibinfo{author}{Lucas, A.}
\newblock \bibinfo{journal}{\bibinfo{title}{Ising formulations of many {NP} problems}}.
\newblock {\emph{\JournalTitle{Frontiers in Physics}}} \textbf{\bibinfo{volume}{2}}, \doiprefix\url{10.3389/fphy.2014.00005} (\bibinfo{year}{2014}).
\newblock \bibinfo{note}{Publisher: Frontiers}.

\bibitem{zhang_mapping_2025}
\bibinfo{author}{Zhang, Z.}
\newblock \bibinfo{title}{Mapping between {Spin}-{Glass} {Three}-{Dimensional} ({3D}) {Ising} {Model} and {Boolean} {Satisfiability} {Problem}}, \doiprefix\url{10.48550/arXiv.2505.18460} (\bibinfo{year}{2025}).
\newblock \bibinfo{note}{ArXiv:2505.18460 [physics]}.

\bibitem{mott_solving_2017}
\bibinfo{author}{Mott, A.}, \bibinfo{author}{Job, J.}, \bibinfo{author}{Vlimant, J.-R.}, \bibinfo{author}{Lidar, D.} \& \bibinfo{author}{Spiropulu, M.}
\newblock \bibinfo{journal}{\bibinfo{title}{Solving a {Higgs} optimization problem with quantum annealing for machine learning}}.
\newblock {\emph{\JournalTitle{Nature}}} \textbf{\bibinfo{volume}{550}}, \bibinfo{pages}{375--379}, \doiprefix\url{10.1038/nature24047} (\bibinfo{year}{2017}).
\newblock \bibinfo{note}{ADS Bibcode: 2017Natur.550..375M}.

\bibitem{king_quantum_2021}
\bibinfo{author}{King, A.~D.} \emph{et~al.}
\newblock \bibinfo{journal}{\bibinfo{title}{Quantum {Annealing} {Simulation} of {Out}-of-{Equilibrium} {Magnetization} in a {Spin}-{Chain} {Compound}}}.
\newblock {\emph{\JournalTitle{PRX Quantum}}} \textbf{\bibinfo{volume}{2}}, \bibinfo{pages}{030317}, \doiprefix\url{10.1103/PRXQuantum.2.030317} (\bibinfo{year}{2021}).
\newblock \bibinfo{note}{Publisher: APS ADS Bibcode: 2021PRXQ....2c0317K}.

\bibitem{abel_quantum-field-theoretic_2021}
\bibinfo{author}{Abel, S.} \& \bibinfo{author}{Spannowsky, M.}
\newblock \bibinfo{journal}{\bibinfo{title}{Quantum-{Field}-{Theoretic} {Simulation} {Platform} for {Observing} the {Fate} of the {False} {Vacuum}}}.
\newblock {\emph{\JournalTitle{PRX Quantum}}} \textbf{\bibinfo{volume}{2}}, \bibinfo{pages}{010349}, \doiprefix\url{10.1103/PRXQuantum.2.010349} (\bibinfo{year}{2021}).
\newblock \bibinfo{note}{Publisher: American Physical Society}.

\bibitem{mohseni_ising_2022}
\bibinfo{author}{Mohseni, N.}, \bibinfo{author}{McMahon, P.~L.} \& \bibinfo{author}{Byrnes, T.}
\newblock \bibinfo{journal}{\bibinfo{title}{Ising machines as hardware solvers of combinatorial optimization problems}}.
\newblock {\emph{\JournalTitle{Nature Reviews Physics}}} \textbf{\bibinfo{volume}{4}}, \bibinfo{pages}{363--379}, \doiprefix\url{10.1038/s42254-022-00440-8} (\bibinfo{year}{2022}).
\newblock \bibinfo{note}{ADS Bibcode: 2022NatRP...4..363M}.

\bibitem{perdomo-ortiz_finding_2012}
\bibinfo{author}{Perdomo-Ortiz, A.}, \bibinfo{author}{Dickson, N.}, \bibinfo{author}{Drew-Brook, M.}, \bibinfo{author}{Rose, G.} \& \bibinfo{author}{Aspuru-Guzik, A.}
\newblock \bibinfo{journal}{\bibinfo{title}{Finding low-energy conformations of lattice protein models by quantum annealing}}.
\newblock {\emph{\JournalTitle{Scientific Reports}}} \textbf{\bibinfo{volume}{2}}, \bibinfo{pages}{571}, \doiprefix\url{10.1038/srep00571} (\bibinfo{year}{2012}).
\newblock \bibinfo{note}{ADS Bibcode: 2012NatSR...2..571P}.

\bibitem{dridi_prime_2017}
\bibinfo{author}{Dridi, R.} \& \bibinfo{author}{Alghassi, H.}
\newblock \bibinfo{journal}{\bibinfo{title}{Prime factorization using quantum annealing and computational algebraic geometry}}.
\newblock {\emph{\JournalTitle{Scientific Reports}}} \textbf{\bibinfo{volume}{7}}, \bibinfo{pages}{43048}, \doiprefix\url{10.1038/srep43048} (\bibinfo{year}{2017}).

\bibitem{jiang_quantum_2018}
\bibinfo{author}{Jiang, S.}, \bibinfo{author}{Britt, K.~A.}, \bibinfo{author}{McCaskey, A.~J.}, \bibinfo{author}{Humble, T.~S.} \& \bibinfo{author}{Kais, S.}
\newblock \bibinfo{journal}{\bibinfo{title}{Quantum {Annealing} for {Prime} {Factorization}}}.
\newblock {\emph{\JournalTitle{Scientific Reports}}} \textbf{\bibinfo{volume}{8}}, \bibinfo{pages}{17667}, \doiprefix\url{10.1038/s41598-018-36058-z} (\bibinfo{year}{2018}).

\bibitem{criado_qade_2022}
\bibinfo{author}{Criado, J.~C.} \& \bibinfo{author}{Spannowsky, M.}
\newblock \bibinfo{title}{Qade: {Solving} {Differential} {Equations} on {Quantum} {Annealers}}, \doiprefix\url{10.48550/arXiv.2204.03657} (\bibinfo{year}{2022}).
\newblock \bibinfo{note}{ArXiv:2204.03657 [quant-ph]}.

\bibitem{paszynski_portfolio_2021}
\bibinfo{author}{Phillipson, F.} \& \bibinfo{author}{Bhatia, H.~S.}
\newblock \bibinfo{title}{Portfolio {Optimisation} {Using} the {D}-{Wave} {Quantum} {Annealer}}.
\newblock In \bibinfo{editor}{Paszynski, M.}, \bibinfo{editor}{Kranzlmüller, D.}, \bibinfo{editor}{Krzhizhanovskaya, V.~V.}, \bibinfo{editor}{Dongarra, J.~J.} \& \bibinfo{editor}{Sloot, P. M.~A.} (eds.) \emph{\bibinfo{booktitle}{Computational {Science} – {ICCS} 2021}}, vol. \bibinfo{volume}{12747}, \bibinfo{pages}{45--59}, \doiprefix\url{10.1007/978-3-030-77980-1_4} (\bibinfo{publisher}{Springer International Publishing}, \bibinfo{address}{Cham}, \bibinfo{year}{2021}).
\newblock \bibinfo{note}{Series Title: Lecture Notes in Computer Science}.

\bibitem{zhang_computational_2025}
\bibinfo{author}{Zhang, H.} \& \bibinfo{author}{Kamenev, A.}
\newblock \bibinfo{title}{On {Computational} {Complexity} of {3D} {Ising} {Spin} {Glass}: {Lessons} from {D}-{Wave} {Annealer}}, \doiprefix\url{10.48550/arXiv.2501.01107} (\bibinfo{year}{2025}).
\newblock \bibinfo{note}{ArXiv:2501.01107 [cond-mat]}.

\bibitem{edwards_theory_1975}
\bibinfo{author}{Edwards, S.~F.} \& \bibinfo{author}{Anderson, P.~W.}
\newblock \bibinfo{journal}{\bibinfo{title}{Theory of spin glasses}}.
\newblock {\emph{\JournalTitle{Journal of Physics F Metal Physics}}} \textbf{\bibinfo{volume}{5}}, \bibinfo{pages}{965--974}, \doiprefix\url{10.1088/0305-4608/5/5/017} (\bibinfo{year}{1975}).
\newblock \bibinfo{note}{Publisher: IOP ADS Bibcode: 1975JPhF....5..965E}.

\bibitem{sherrington_solvable_1975}
\bibinfo{author}{Sherrington, D.} \& \bibinfo{author}{Kirkpatrick, S.}
\newblock \bibinfo{journal}{\bibinfo{title}{Solvable {Model} of a {Spin}-{Glass}}}.
\newblock {\emph{\JournalTitle{Physical Review Letters}}} \textbf{\bibinfo{volume}{35}}, \bibinfo{pages}{1792--1796}, \doiprefix\url{10.1103/PhysRevLett.35.1792} (\bibinfo{year}{1975}).
\newblock \bibinfo{note}{Publisher: American Physical Society}.

\bibitem{young_spin_2002}
\bibinfo{author}{Young, A.}
\newblock \bibinfo{journal}{\bibinfo{title}{Spin glasses: a computational challenge for the 21st century}}.
\newblock {\emph{\JournalTitle{Computer Physics Communications}}} \textbf{\bibinfo{volume}{146}}, \bibinfo{pages}{107--112}, \doiprefix\url{10.1016/S0010-4655(02)00441-1} (\bibinfo{year}{2002}).

\bibitem{boettcher_physics_2024}
\bibinfo{author}{Boettcher, S.}
\newblock \bibinfo{journal}{\bibinfo{title}{Physics of the {Edwards}–{Anderson} spin glass in dimensions d = 3, … ,8 from heuristic ground state optimization}}.
\newblock {\emph{\JournalTitle{Frontiers in Physics}}} \textbf{\bibinfo{volume}{12}}, \bibinfo{pages}{1466987}, \doiprefix\url{10.3389/fphy.2024.1466987} (\bibinfo{year}{2024}).

\bibitem{istrail_statistical_2000}
\bibinfo{author}{Istrail, S.}
\newblock \bibinfo{title}{Statistical mechanics, three-dimensionality and {NP}-completeness: {I}. {Universality} of intracatability for the partition function of the {Ising} model across non-planar surfaces (extended abstract)}.
\newblock In \emph{\bibinfo{booktitle}{Proceedings of the thirty-second annual {ACM} symposium on {Theory} of computing}}, {STOC} '00, \bibinfo{pages}{87--96}, \doiprefix\url{10.1145/335305.335316} (\bibinfo{publisher}{Association for Computing Machinery}, \bibinfo{address}{New York, NY, USA}, \bibinfo{year}{2000}).

\bibitem{hartmann_ground_2011}
\bibinfo{author}{Hartmann, A.~K.}
\newblock \bibinfo{journal}{\bibinfo{title}{Ground {States} of {Two}-{Dimensional} {Ising} {Spin} {Glasses}: {Fast} {Algorithms}, {Recent} {Developments} and a {Ferromagnet}-{Spin} {Glass} {Mixture}}}.
\newblock {\emph{\JournalTitle{Journal of Statistical Physics}}} \textbf{\bibinfo{volume}{144}}, \bibinfo{pages}{519--540}, \doiprefix\url{10.1007/s10955-011-0272-1} (\bibinfo{year}{2011}).

\bibitem{garey_computers_1979}
\bibinfo{author}{Garey, M.~R.} \& \bibinfo{author}{Johnson, D.~S.}
\newblock \emph{\bibinfo{title}{Computers and intractability: a guide to the theory of {NP}-completeness}}.
\newblock A {Series} of books in the mathematical sciences (\bibinfo{publisher}{W. H. Freeman}, \bibinfo{address}{San Francisco}, \bibinfo{year}{1979}).

\bibitem{arora_computational_2009}
\bibinfo{author}{Arora, S.} \& \bibinfo{author}{Barak, B.}
\newblock \emph{\bibinfo{title}{Computational complexity: a modern approach}} (\bibinfo{publisher}{Cambridge University Press}, \bibinfo{address}{Cambridge ; New York}, \bibinfo{year}{2009}).

\bibitem{hastings_monte_1970}
\bibinfo{author}{Hastings, W.~K.}
\newblock \bibinfo{journal}{\bibinfo{title}{Monte {Carlo} {Sampling} {Methods} using {Markov} {Chains} and their {Applications}}}.
\newblock {\emph{\JournalTitle{Biometrika}}} \textbf{\bibinfo{volume}{57}}, \bibinfo{pages}{97--109}, \doiprefix\url{10.1093/biomet/57.1.97} (\bibinfo{year}{1970}).
\newblock \bibinfo{note}{ADS Bibcode: 1970Bimka..57...97H}.

\bibitem{kirkpatrick_optimization_1983}
\bibinfo{author}{Kirkpatrick, S.}, \bibinfo{author}{Gelatt, C.~D.} \& \bibinfo{author}{Vecchi, M.~P.}
\newblock \bibinfo{journal}{\bibinfo{title}{Optimization by {Simulated} {Annealing}}}.
\newblock {\emph{\JournalTitle{Science}}} \textbf{\bibinfo{volume}{220}}, \bibinfo{pages}{671--680}, \doiprefix\url{10.1126/science.220.4598.671} (\bibinfo{year}{1983}).
\newblock \bibinfo{note}{ADS Bibcode: 1983Sci...220..671K}.

\bibitem{zhang_computational_2020}
\bibinfo{author}{Zhang, Z.}
\newblock \bibinfo{journal}{\bibinfo{title}{Computational complexity of spin-glass three-dimensional ({3D}) {Ising} model}}.
\newblock {\emph{\JournalTitle{Journal of Materials Science \& Technology}}} \textbf{\bibinfo{volume}{44}}, \bibinfo{pages}{116--120}, \doiprefix\url{10.1016/j.jmst.2019.12.009} (\bibinfo{year}{2020}).

\bibitem{impagliazzo_complexity_1999}
\bibinfo{author}{Impagliazzo, R.} \& \bibinfo{author}{Paturi, R.}
\newblock \bibinfo{title}{Complexity of k-{SAT}}.
\newblock In \emph{\bibinfo{booktitle}{Proceedings. {Fourteenth} {Annual} {IEEE} {Conf}. {Comp}. {Compl}.}}, \bibinfo{pages}{237--240}, \doiprefix\url{10.1109/CCC.1999.766282} (\bibinfo{publisher}{IEEE Comput. Soc}, \bibinfo{address}{Atlanta, GA, USA}, \bibinfo{year}{1999}).

\bibitem{johnson_quantum_2011}
\bibinfo{author}{Johnson, M.~W.} \emph{et~al.}
\newblock \bibinfo{journal}{\bibinfo{title}{Quantum annealing with manufactured spins}}.
\newblock {\emph{\JournalTitle{Nature}}} \textbf{\bibinfo{volume}{473}}, \bibinfo{pages}{194--198}, \doiprefix\url{10.1038/nature10012} (\bibinfo{year}{2011}).

\bibitem{dickson_thermally_2013}
\bibinfo{author}{Dickson, N.~G.} \emph{et~al.}
\newblock \bibinfo{journal}{\bibinfo{title}{Thermally assisted quantum annealing of a 16-qubit problem}}.
\newblock {\emph{\JournalTitle{Nature Communications}}} \textbf{\bibinfo{volume}{4}}, \bibinfo{pages}{1903}, \doiprefix\url{10.1038/ncomms2920} (\bibinfo{year}{2013}).
\newblock \bibinfo{note}{ADS Bibcode: 2013NatCo...4.1903D}.

\bibitem{ohkuwa_reverse_2018}
\bibinfo{author}{Ohkuwa, M.}, \bibinfo{author}{Nishimori, H.} \& \bibinfo{author}{Lidar, D.~A.}
\newblock \bibinfo{journal}{\bibinfo{title}{Reverse annealing for the fully connected p -spin model}}.
\newblock {\emph{\JournalTitle{Physical Review A}}} \textbf{\bibinfo{volume}{98}}, \bibinfo{pages}{022314}, \doiprefix\url{10.1103/PhysRevA.98.022314} (\bibinfo{year}{2018}).

\bibitem{cao_speedup_2021}
\bibinfo{author}{Cao, C.}, \bibinfo{author}{Xue, J.}, \bibinfo{author}{Shannon, N.} \& \bibinfo{author}{Joynt, R.}
\newblock \bibinfo{journal}{\bibinfo{title}{Speedup of the quantum adiabatic algorithm using delocalization catalysis}}.
\newblock {\emph{\JournalTitle{Physical Review Research}}} \textbf{\bibinfo{volume}{3}}, \bibinfo{pages}{013092}, \doiprefix\url{10.1103/PhysRevResearch.3.013092} (\bibinfo{year}{2021}).

\bibitem{king_coherent_2022}
\bibinfo{author}{King, A.~D.} \emph{et~al.}
\newblock \bibinfo{journal}{\bibinfo{title}{Coherent quantum annealing in a programmable 2,000 qubit {Ising} chain}}.
\newblock {\emph{\JournalTitle{Nature Physics}}} \textbf{\bibinfo{volume}{18}}, \bibinfo{pages}{1324--1328}, \doiprefix\url{10.1038/s41567-022-01741-6} (\bibinfo{year}{2022}).

\bibitem{king_quantum_2023}
\bibinfo{author}{King, A.~D.} \emph{et~al.}
\newblock \bibinfo{journal}{\bibinfo{title}{Quantum critical dynamics in a 5,000-qubit programmable spin glass}}.
\newblock {\emph{\JournalTitle{Nature}}} \textbf{\bibinfo{volume}{617}}, \bibinfo{pages}{61--66}, \doiprefix\url{10.1038/s41586-023-05867-2} (\bibinfo{year}{2023}).

\bibitem{bernaschi_quantum_2024}
\bibinfo{author}{Bernaschi, M.}, \bibinfo{author}{González-Adalid~Pemartín, I.}, \bibinfo{author}{Martín-Mayor, V.} \& \bibinfo{author}{Parisi, G.}
\newblock \bibinfo{journal}{\bibinfo{title}{The quantum transition of the two-dimensional {Ising} spin glass}}.
\newblock {\emph{\JournalTitle{Nature}}} \textbf{\bibinfo{volume}{631}}, \bibinfo{pages}{749--754}, \doiprefix\url{10.1038/s41586-024-07647-y} (\bibinfo{year}{2024}).

\bibitem{zbinden_embedding_2020}
\bibinfo{author}{Zbinden, S.}, \bibinfo{author}{Bärtschi, A.}, \bibinfo{author}{Djidjev, H.} \& \bibinfo{author}{Eidenbenz, S.}
\newblock \bibinfo{journal}{\bibinfo{title}{Embedding {Algorithms} for {Quantum} {Annealers} with {Chimera} and {Pegasus} {Connection} {Topologies}}}.
\newblock {\emph{\JournalTitle{High Performance Computing}}} \textbf{\bibinfo{volume}{12151}}, \bibinfo{pages}{187--206}, \doiprefix\url{10.1007/978-3-030-50743-5_10} (\bibinfo{year}{2020}).

\bibitem{hamerly_experimental_2019}
\bibinfo{author}{Hamerly, R.} \emph{et~al.}
\newblock \bibinfo{journal}{\bibinfo{title}{Experimental investigation of performance differences between coherent {Ising} machines and a quantum annealer}}.
\newblock {\emph{\JournalTitle{Science Advances}}} \textbf{\bibinfo{volume}{5}}, \bibinfo{pages}{eaau0823}, \doiprefix\url{10.1126/sciadv.aau0823} (\bibinfo{year}{2019}).

\bibitem{zeng_performance_2024}
\bibinfo{author}{Zeng, Q.-G.} \emph{et~al.}
\newblock \bibinfo{journal}{\bibinfo{title}{Performance of quantum annealing inspired algorithms for combinatorial optimization problems}}.
\newblock {\emph{\JournalTitle{Communications Physics}}} \textbf{\bibinfo{volume}{7}}, \bibinfo{pages}{249}, \doiprefix\url{10.1038/s42005-024-01705-7} (\bibinfo{year}{2024}).
\newblock \bibinfo{note}{Publisher: Nature Publishing Group}.

\bibitem{bunyk_architectural_2014}
\bibinfo{author}{Bunyk, P.~I.} \emph{et~al.}
\newblock \bibinfo{journal}{\bibinfo{title}{Architectural {Considerations} in the {Design} of a {Superconducting} {Quantum} {Annealing} {Processor}}}.
\newblock {\emph{\JournalTitle{IEEE Transactions on Applied Superconductivity}}} \textbf{\bibinfo{volume}{24}}, \bibinfo{pages}{1--10}, \doiprefix\url{10.1109/TASC.2014.2318294} (\bibinfo{year}{2014}).

\bibitem{boixo_evidence_2014}
\bibinfo{author}{Boixo, S.} \emph{et~al.}
\newblock \bibinfo{journal}{\bibinfo{title}{Evidence for quantum annealing with more than one hundred qubits}}.
\newblock {\emph{\JournalTitle{Nature Physics}}} \textbf{\bibinfo{volume}{10}}, \bibinfo{pages}{218--224}, \doiprefix\url{10.1038/nphys2900} (\bibinfo{year}{2014}).
\newblock \bibinfo{note}{Publisher: Nature Publishing Group}.

\bibitem{boev_quantum-inspired_2023}
\bibinfo{author}{Boev, A.~S.} \emph{et~al.}
\newblock \bibinfo{journal}{\bibinfo{title}{Quantum-inspired optimization for wavelength assignment}}.
\newblock {\emph{\JournalTitle{Frontiers in Physics}}} \textbf{\bibinfo{volume}{10}}, \bibinfo{pages}{1092065}, \doiprefix\url{10.3389/fphy.2022.1092065} (\bibinfo{year}{2023}).

\bibitem{okada_improving_2019}
\bibinfo{author}{Okada, S.}, \bibinfo{author}{Ohzeki, M.}, \bibinfo{author}{Terabe, M.} \& \bibinfo{author}{Taguchi, S.}
\newblock \bibinfo{journal}{\bibinfo{title}{Improving solutions by embedding larger subproblems in a {D}-{Wave} quantum annealer}}.
\newblock {\emph{\JournalTitle{Scientific Reports}}} \textbf{\bibinfo{volume}{9}}, \bibinfo{pages}{2098}, \doiprefix\url{10.1038/s41598-018-38388-4} (\bibinfo{year}{2019}).

\bibitem{king_beyond-classical_2025}
\bibinfo{author}{King, A.~D.} \emph{et~al.}
\newblock \bibinfo{journal}{\bibinfo{title}{Beyond-classical computation in quantum simulation}}.
\newblock {\emph{\JournalTitle{Science}}} \textbf{\bibinfo{volume}{388}}, \bibinfo{pages}{199--204}, \doiprefix\url{10.1126/science.ado6285} (\bibinfo{year}{2025}).
\newblock \bibinfo{note}{Publisher: American Association for the Advancement of Science}.

\bibitem{tindall_dynamics_2025}
\bibinfo{author}{Tindall, J.}, \bibinfo{author}{Mello, A.}, \bibinfo{author}{Fishman, M.}, \bibinfo{author}{Stoudenmire, M.} \& \bibinfo{author}{Sels, D.}
\newblock \bibinfo{title}{Dynamics of disordered quantum systems with two- and three-dimensional tensor networks}, \doiprefix\url{10.48550/arXiv.2503.05693} (\bibinfo{year}{2025}).
\newblock \bibinfo{note}{ArXiv:2503.05693 [quant-ph]}.

\bibitem{tindall_efficient_2024}
\bibinfo{author}{Tindall, J.}, \bibinfo{author}{Fishman, M.}, \bibinfo{author}{Stoudenmire, E.~M.} \& \bibinfo{author}{Sels, D.}
\newblock \bibinfo{journal}{\bibinfo{title}{Efficient {Tensor} {Network} {Simulation} of {IBM}’s {Eagle} {Kicked} {Ising} {Experiment}}}.
\newblock {\emph{\JournalTitle{PRX Quantum}}} \textbf{\bibinfo{volume}{5}}, \bibinfo{pages}{010308}, \doiprefix\url{10.1103/PRXQuantum.5.010308} (\bibinfo{year}{2024}).

\bibitem{rudolph_simulating_2025}
\bibinfo{author}{Rudolph, M.~S.} \& \bibinfo{author}{Tindall, J.}
\newblock \bibinfo{title}{Simulating and {Sampling} from {Quantum} {Circuits} with {2D} {Tensor} {Networks}}, \doiprefix\url{10.48550/arXiv.2507.11424} (\bibinfo{year}{2025}).
\newblock \bibinfo{note}{ArXiv:2507.11424 [quant-ph]}.

\bibitem{begusic_fast_2024}
\bibinfo{author}{Begušić, T.}, \bibinfo{author}{Gray, J.} \& \bibinfo{author}{Chan, G. K.-L.}
\newblock \bibinfo{journal}{\bibinfo{title}{Fast and converged classical simulations of evidence for the utility of quantum computing before fault tolerance}}.
\newblock {\emph{\JournalTitle{Science Advances}}} \textbf{\bibinfo{volume}{10}}, \bibinfo{pages}{eadk4321}, \doiprefix\url{10.1126/sciadv.adk4321} (\bibinfo{year}{2024}).

\bibitem{garey_simplified_1976}
\bibinfo{author}{Garey, M.~R.}, \bibinfo{author}{Johnson, D.~S.} \& \bibinfo{author}{Stockmeyer, L.}
\newblock \bibinfo{journal}{\bibinfo{title}{Some simplified \textit{{NP}}-complete graph problems}}.
\newblock {\emph{\JournalTitle{Theoretical Computer Science}}} \textbf{\bibinfo{volume}{1}}, \bibinfo{pages}{237--267}, \doiprefix\url{10.1016/0304-3975(76)90059-1} (\bibinfo{year}{1976}).

\bibitem{williams_new_2004}
\bibinfo{author}{Williams, R.}
\newblock \bibinfo{title}{A {New} {Algorithm} for {Optimal} {Constraint} {Satisfaction} and {Its} {Implications}}.
\newblock In \bibinfo{editor}{Díaz, J.}, \bibinfo{editor}{Karhumäki, J.}, \bibinfo{editor}{Lepistö, A.} \& \bibinfo{editor}{Sannella, D.} (eds.) \emph{\bibinfo{booktitle}{Automata, {Languages} and {Programming}}}, \bibinfo{pages}{1227--1237}, \doiprefix\url{10.1007/978-3-540-27836-8_101} (\bibinfo{publisher}{Springer}, \bibinfo{address}{Berlin, Heidelberg}, \bibinfo{year}{2004}).

\bibitem{lalovic_exact_2024}
\bibinfo{author}{Lalovic, M.}
\newblock \bibinfo{title}{Exact {Algorithms} for {MaxCut} on {Split} {Graphs}}, \doiprefix\url{10.48550/arXiv.2405.20599} (\bibinfo{year}{2024}).
\newblock \bibinfo{note}{ArXiv:2405.20599 [cs] version: 1}.

\bibitem{woeginger_open_2008}
\bibinfo{author}{Woeginger, G.~J.}
\newblock \bibinfo{journal}{\bibinfo{title}{Open problems around exact algorithms}}.
\newblock {\emph{\JournalTitle{Discrete Applied Mathematics}}} \textbf{\bibinfo{volume}{156}}, \bibinfo{pages}{397--405}, \doiprefix\url{10.1016/j.dam.2007.03.023} (\bibinfo{year}{2008}).

\bibitem{leleu_scaling_2021}
\bibinfo{author}{Leleu, T.} \emph{et~al.}
\newblock \bibinfo{title}{Scaling advantage of nonrelaxational dynamics for high-performance combinatorial optimization}, \doiprefix\url{10.48550/arXiv.2009.04084} (\bibinfo{year}{2021}).
\newblock \bibinfo{note}{ArXiv:2009.04084 [physics]}.

\bibitem{kowalsky_3-regular_2022}
\bibinfo{author}{Kowalsky, M.}, \bibinfo{author}{Albash, T.}, \bibinfo{author}{Hen, I.} \& \bibinfo{author}{Lidar, D.~A.}
\newblock \bibinfo{journal}{\bibinfo{title}{3-{Regular} 3-{XORSAT} {Planted} {Solutions} {Benchmark} of {Classical} and {Quantum} {Heuristic} {Optimizers}}}.
\newblock {\emph{\JournalTitle{Quantum Science and Technology}}} \textbf{\bibinfo{volume}{7}}, \bibinfo{pages}{025008}, \doiprefix\url{10.1088/2058-9565/ac4d1b} (\bibinfo{year}{2022}).
\newblock \bibinfo{note}{ArXiv:2103.08464 [quant-ph]}.

\bibitem{mandra_deceptive_2018}
\bibinfo{author}{Mandrà, S.} \& \bibinfo{author}{Katzgraber, H.~G.}
\newblock \bibinfo{journal}{\bibinfo{title}{A deceptive step towards quantum speedup detection}}.
\newblock {\emph{\JournalTitle{Quantum Science and Technology}}} \textbf{\bibinfo{volume}{3}}, \bibinfo{pages}{04LT01}, \doiprefix\url{10.1088/2058-9565/aac8b2} (\bibinfo{year}{2018}).
\newblock \bibinfo{note}{ArXiv:1711.01368 [quant-ph]}.

\bibitem{aramon_physics-inspired_2019}
\bibinfo{author}{Aramon, M.} \emph{et~al.}
\newblock \bibinfo{journal}{\bibinfo{title}{Physics-{Inspired} {Optimization} for {Quadratic} {Unconstrained} {Problems} {Using} a {Digital} {Annealer}}}.
\newblock {\emph{\JournalTitle{Frontiers in Physics}}} \textbf{\bibinfo{volume}{7}}, \bibinfo{pages}{48}, \doiprefix\url{10.3389/fphy.2019.00048} (\bibinfo{year}{2019}).
\newblock \bibinfo{note}{ArXiv:1806.08815 [physics]}.

\bibitem{cai_power-efficient_2020}
\bibinfo{author}{Cai, F.} \emph{et~al.}
\newblock \bibinfo{journal}{\bibinfo{title}{Power-efficient combinatorial optimization using intrinsic noise in memristor {Hopfield} neural networks}}.
\newblock {\emph{\JournalTitle{Nature Electronics}}} \textbf{\bibinfo{volume}{3}}, \bibinfo{pages}{409--418}, \doiprefix\url{10.1038/s41928-020-0436-6} (\bibinfo{year}{2020}).
\newblock \bibinfo{note}{Publisher: Nature Publishing Group}.

\bibitem{patel_ising_2020}
\bibinfo{author}{Patel, S.}, \bibinfo{author}{Chen, L.}, \bibinfo{author}{Canoza, P.} \& \bibinfo{author}{Salahuddin, S.}
\newblock \bibinfo{title}{Ising {Model} {Optimization} {Problems} on a {FPGA} {Accelerated} {Restricted} {Boltzmann} {Machine}}, \doiprefix\url{10.48550/arXiv.2008.04436} (\bibinfo{year}{2020}).
\newblock \bibinfo{note}{ArXiv:2008.04436 [cs]}.

\bibitem{albash_demonstration_2018}
\bibinfo{author}{Albash, T.} \& \bibinfo{author}{Lidar, D.~A.}
\newblock \bibinfo{journal}{\bibinfo{title}{Demonstration of a {Scaling} {Advantage} for a {Quantum} {Annealer} over {Simulated} {Annealing}}}.
\newblock {\emph{\JournalTitle{Physical Review X}}} \textbf{\bibinfo{volume}{8}}, \bibinfo{pages}{031016}, \doiprefix\url{10.1103/PhysRevX.8.031016} (\bibinfo{year}{2018}).

\bibitem{blais_operation_2000}
\bibinfo{author}{Blais, A.} \& \bibinfo{author}{Zagoskin, A.~M.}
\newblock \bibinfo{journal}{\bibinfo{title}{Operation of universal gates in a solid-state quantum computer based on clean {Josephson} junctions between \textbf{ \textit{d} } -wave superconductors}}.
\newblock {\emph{\JournalTitle{Physical Review A}}} \textbf{\bibinfo{volume}{61}}, \bibinfo{pages}{042308}, \doiprefix\url{10.1103/PhysRevA.61.042308} (\bibinfo{year}{2000}).

\bibitem{inagaki_coherent_2016}
\bibinfo{author}{Inagaki, T.} \emph{et~al.}
\newblock \bibinfo{journal}{\bibinfo{title}{A coherent {Ising} machine for 2000-node optimization problems}}.
\newblock {\emph{\JournalTitle{Science}}} \textbf{\bibinfo{volume}{354}}, \bibinfo{pages}{603--606}, \doiprefix\url{10.1126/science.aah4243} (\bibinfo{year}{2016}).

\bibitem{mcmahon_fully_2016}
\bibinfo{author}{McMahon, P.~L.} \emph{et~al.}
\newblock \bibinfo{journal}{\bibinfo{title}{A fully programmable 100-spin coherent {Ising} machine with all-to-all connections}}.
\newblock {\emph{\JournalTitle{Science}}} \textbf{\bibinfo{volume}{354}}, \bibinfo{pages}{614--617}, \doiprefix\url{10.1126/science.aah5178} (\bibinfo{year}{2016}).

\bibitem{camsari_stochastic_2017}
\bibinfo{author}{Camsari, K.~Y.}, \bibinfo{author}{Faria, R.}, \bibinfo{author}{Sutton, B.~M.} \& \bibinfo{author}{Datta, S.}
\newblock \bibinfo{journal}{\bibinfo{title}{Stochastic \$p\$-{Bits} for {Invertible} {Logic}}}.
\newblock {\emph{\JournalTitle{Physical Review X}}} \textbf{\bibinfo{volume}{7}}, \bibinfo{pages}{031014}, \doiprefix\url{10.1103/PhysRevX.7.031014} (\bibinfo{year}{2017}).
\newblock \bibinfo{note}{Publisher: American Physical Society}.

\bibitem{chowdhury_full-stack_2023}
\bibinfo{author}{Chowdhury, S.} \emph{et~al.}
\newblock \bibinfo{journal}{\bibinfo{title}{A {Full}-{Stack} {View} of {Probabilistic} {Computing} {With} p-{Bits}: {Devices}, {Architectures}, and {Algorithms}}}.
\newblock {\emph{\JournalTitle{IEEE Journal on Exploratory Solid-State Computational Devices and Circuits}}} \textbf{\bibinfo{volume}{9}}, \bibinfo{pages}{1--11}, \doiprefix\url{10.1109/JXCDC.2023.3256981} (\bibinfo{year}{2023}).
\newblock \bibinfo{note}{Conference Name: IEEE Journal on Exploratory Solid-State Computational Devices and Circuits}.

\bibitem{mascagni_simulated_1990}
\bibinfo{author}{Mascagni, M.}, \bibinfo{author}{Aart, E.} \& \bibinfo{author}{Korst, J.}
\newblock \bibinfo{title}{Simulated {Annealing} and {Boltzmann} {Machines}: {A} {Stochastic} {Approach} to {Combinatorial} {Optimization} and {Neural} {Computing}.}
\newblock In \emph{\bibinfo{booktitle}{Mathematics of {Computation}}}, vol.~\bibinfo{volume}{55}, \bibinfo{pages}{393}, \doiprefix\url{10.2307/2008816} (\bibinfo{year}{1990}).
\newblock \bibinfo{note}{ISSN: 00255718 Issue: 191 Journal Abbreviation: Mathematics of Computation}.

\bibitem{borowiecki_computational_2018}
\bibinfo{author}{Borowiecki, P.}
\newblock \bibinfo{journal}{\bibinfo{title}{Computational aspects of greedy partitioning of graphs}}.
\newblock {\emph{\JournalTitle{Journal of Combinatorial Optimization}}} \textbf{\bibinfo{volume}{35}}, \bibinfo{pages}{641--665}, \doiprefix\url{10.1007/s10878-017-0185-2} (\bibinfo{year}{2018}).

\bibitem{dunning_what_2018}
\bibinfo{author}{Dunning, I.}, \bibinfo{author}{Gupta, S.} \& \bibinfo{author}{Silberholz, J.}
\newblock \bibinfo{journal}{\bibinfo{title}{What {Works} {Best} {When}? {A} {Systematic} {Evaluation} of {Heuristics} for {Max}-{Cut} and {QUBO}}}.
\newblock {\emph{\JournalTitle{INFORMS Journal on Computing}}} \textbf{\bibinfo{volume}{30}}, \bibinfo{pages}{608--624}, \doiprefix\url{10.1287/ijoc.2017.0798} (\bibinfo{year}{2018}).
\newblock \bibinfo{note}{Publisher: INFORMS}.

\bibitem{viering_shape_2023}
\bibinfo{author}{Viering, T.} \& \bibinfo{author}{Marco, L.}
\newblock \bibinfo{journal}{\bibinfo{title}{The {Shape} of {Learning} {Curves}: {A} {Review}}}.
\newblock {\emph{\JournalTitle{IEEE Transactions on Pattern Analysis and Machine Intelligence}}}  (\bibinfo{year}{2023}).

\bibitem{krzakala_spin_2000}
\bibinfo{author}{Krzakala, F.}
\newblock \bibinfo{journal}{\bibinfo{title}{Spin and {Link} {Overlaps} in {Three}-{Dimensional} {Spin} {Glasses}}}.
\newblock {\emph{\JournalTitle{Physical Review Letters}}} \textbf{\bibinfo{volume}{85}}, \bibinfo{pages}{3013--3016}, \doiprefix\url{10.1103/PhysRevLett.85.3013} (\bibinfo{year}{2000}).

\bibitem{palassini_nature_2000}
\bibinfo{author}{Palassini, M.}
\newblock \bibinfo{journal}{\bibinfo{title}{Nature of the {Spin} {Glass} {State}}}.
\newblock {\emph{\JournalTitle{Physical Review Letters}}} \textbf{\bibinfo{volume}{85}}, \bibinfo{pages}{3017--3020}, \doiprefix\url{10.1103/PhysRevLett.85.3017} (\bibinfo{year}{2000}).

\bibitem{boothby_next-generation_2020}
\bibinfo{author}{Boothby, K.}, \bibinfo{author}{Bunyk, P.}, \bibinfo{author}{Raymond, J.} \& \bibinfo{author}{Roy, A.}
\newblock \bibinfo{title}{Next-{Generation} {Topology} of {D}-{Wave} {Quantum} {Processors}}, \doiprefix\url{10.48550/arXiv.2003.00133} (\bibinfo{year}{2020}).
\newblock \bibinfo{note}{ArXiv:2003.00133 [quant-ph]}.

\bibitem{al-hasso_probabilistic_2025}
\bibinfo{author}{Al-Hasso, M.~O.} \& \bibinfo{author}{von~der Leyen, M.}
\newblock \bibinfo{title}{A {Probabilistic} {Computing} {Approach} to the {Closest} {Vector} {Problem} for {Lattice}-{Based} {Factoring}}, \doiprefix\url{https://doi.org/10.48550/arXiv.2510.19390} (\bibinfo{year}{2025}).
\newblock \bibinfo{note}{ArXiv:2510.19390v1 [cs.CR]}.

\bibitem{mezard_information_2009}
\bibinfo{author}{Mézard, M.}
\newblock \emph{\bibinfo{title}{Information, {Physics}, and {Computation}}}.
\newblock Oxford {Graduate} {Texts} (\bibinfo{publisher}{Oxford University Press, Incorporated}, \bibinfo{address}{Oxford}, \bibinfo{year}{2009}).

\bibitem{palassini_low-energy_2003}
\bibinfo{author}{Palassini, M.}, \bibinfo{author}{Liers, F.}, \bibinfo{author}{Juenger, M.} \& \bibinfo{author}{Young, A.~P.}
\newblock \bibinfo{journal}{\bibinfo{title}{Low-energy excitations in spin glasses from exact ground states}}.
\newblock {\emph{\JournalTitle{Physical Review B}}} \textbf{\bibinfo{volume}{68}}, \bibinfo{pages}{064413}, \doiprefix\url{10.1103/PhysRevB.68.064413} (\bibinfo{year}{2003}).

\bibitem{searle_virtually_2025}
\bibinfo{author}{Searle, A.~J.} \emph{et~al.}
\newblock \bibinfo{title}{A {Virtually} {Connected} {Probabilistic} {Computer} as a {Solver} for {Higher}-{Order}, {Densely} {Connected}, or {Reconfigurable} {Combinatorial} {Optimisation} {Problems}} (\bibinfo{year}{2025}).
\newblock \bibinfo{note}{Published: preprint}.

\bibitem{aadit_massively_2022}
\bibinfo{author}{Aadit, N.~A.} \emph{et~al.}
\newblock \bibinfo{journal}{\bibinfo{title}{Massively parallel probabilistic computing with sparse {Ising} machines}}.
\newblock {\emph{\JournalTitle{Nature Electronics}}} \textbf{\bibinfo{volume}{5}}, \bibinfo{pages}{460--468}, \doiprefix\url{10.1038/s41928-022-00774-2} (\bibinfo{year}{2022}).
\newblock \bibinfo{note}{Publisher: Nature Publishing Group}.

\bibitem{sajeeb_scalable_2025}
\bibinfo{author}{Sajeeb, M. M.~H.} \emph{et~al.}
\newblock \bibinfo{journal}{\bibinfo{title}{Scalable {Connectivity} for {Ising} {Machines}: {Dense} to {Sparse}}}.
\newblock {\emph{\JournalTitle{Physical Review Applied}}} \textbf{\bibinfo{volume}{24}}, \bibinfo{pages}{014005}, \doiprefix\url{10.1103/kx8m-5h3h} (\bibinfo{year}{2025}).
\newblock \bibinfo{note}{ArXiv:2503.01177 [cs]}.

\bibitem{nikhar_all--all_2024}
\bibinfo{author}{Nikhar, S.}, \bibinfo{author}{Kannan, S.}, \bibinfo{author}{Aadit, N.~A.}, \bibinfo{author}{Chowdhury, S.} \& \bibinfo{author}{Camsari, K.~Y.}
\newblock \bibinfo{journal}{\bibinfo{title}{All-to-all reconfigurability with sparse and higher-order {Ising} machines}}.
\newblock {\emph{\JournalTitle{Nature Communications}}} \textbf{\bibinfo{volume}{15}}, \bibinfo{pages}{8977}, \doiprefix\url{10.1038/s41467-024-53270-w} (\bibinfo{year}{2024}).

\bibitem{chowdhury_pushing_2025}
\bibinfo{author}{Chowdhury, S.} \emph{et~al.}
\newblock \bibinfo{title}{Pushing the {Boundary} of {Quantum} {Advantage} in {Hard} {Combinatorial} {Optimization} with {Probabilistic} {Computers}}, \doiprefix\url{10.48550/arXiv.2503.10302} (\bibinfo{year}{2025}).
\newblock \bibinfo{note}{ArXiv:2503.10302 [quant-ph]}.

\end{thebibliography}
